\documentclass{aa}  
\usepackage{siunitx}
\usepackage{graphicx}
\usepackage{txfonts}
\usepackage{natbib}
\usepackage[
        colorlinks=true,
        citecolor=blue,
        linkcolor=blue]{hyperref}
\begin{document}

   \title{A new scenario for magnetar formation: Tayler-Spruit dynamo in a proto-neutron star spun up by fallback}

   \author{P. Barrère\thanks{\email{paul.barrere@cea.fr}}\inst{1} \and J. Guilet\inst{1} \and A. Reboul-Salze\inst{2} \and R. Raynaud\inst{3} \and H.-T. Janka\inst{4}  }

   \institute{\inst{1}Université Paris-Saclay, Université Paris Cité, CEA, CNRS, AIM, 91191, Gif-sur-Yvette, France\\
   \inst{2}Max Planck Institute for Gravitational Physics (Albert Einstein Institute), D-14476 Potsdam, Germany\\
   \inst{3}Universit\'e Paris Cit\'e, Universit\'e Paris-Saclay, CNRS, CEA, AIM, F-91191 Gif-sur-Yvette, France\\
\inst{4}Max Planck Institute for Astrophysics, Karl-Schwarzschild-Str. 1, 85748 Garching, Germany             }

   \date{Received 2 June 2022 / Accepted 11 October 2022}
 
  \abstract{Magnetars are isolated young neutron stars characterised by the most intense magnetic fields known in the Universe, which power a wide variety of high-energy emissions from giant flares to fast radio bursts. The origin of their magnetic field is still a challenging question. In situ magnetic field amplification by dynamo action could potentially generate ultra-strong magnetic fields in fast-rotating progenitors. However, it is unclear whether the fraction of progenitors harbouring fast core rotation is sufficient to explain the entire magnetar population. To address this point, we propose a new scenario for magnetar formation involving a slowly rotating progenitor, in which a slow-rotating proto-neutron star is spun up by the supernova fallback. We argue that this can trigger the development of the Tayler-Spruit dynamo while other dynamo processes are disfavoured. Using the findings of previous studies of this dynamo and simulation results characterising the supernova fallback, we derive equations modelling the coupled evolution of the proto-neutron star rotation and magnetic field. Their time integration for different accreted masses is successfully compared  with analytical estimates of the amplification timescales and saturation value of the magnetic field. We find that the magnetic field is amplified within 20 to $\SI{40}{s}$ after the core bounce, and that the radial magnetic field saturates at intensities between $\sim\SI{e13}{}$ and $\SI{e15}{G}$, therefore spanning the full range of a magnetar's dipolar magnetic fields. The toroidal magnetic field is predicted to be a factor of 10 to 100 times stronger, lying between $\sim\SI{e15}{}$ and $\SI{3e16}{G}$. We also compare the saturation mechanisms proposed respectively by H.C.~Spruit and J.~Fuller, showing that magnetar-like magnetic fields can be generated for a neutron star spun up to rotation periods of $\lesssim\SI{8}{ms}$ and $\lesssim\SI{28}{ms}$, corresponding to accreted masses of $\gtrsim\SI{4e-2}{M_{\odot}}$ and $\gtrsim\SI{1.1e-2}{M_{\odot}}$, respectively. Therefore, our results suggest that magnetars can be formed from slow-rotating progenitors for accreted masses compatible with recent supernova simulations and leading to plausible initial rotation periods of the proto-neutron star.}


   \maketitle
   

\section{Introduction}
Magnetars represent two classes of isolated young neutron stars whose emission is powered by their ultrastrong magnetic field: anomalous X-ray pulsars and soft gamma repeaters. They feature a large spectrum of activity from short bursts \citep{gotz2006,coti2018,coti2021} to giant flares \citep{evans1980,hurley1999,hurley2005,svinkin2021}, whose signal contains quasi-periodic oscillations \citep{israel2005,strohmayer2005,gabler2018,roberts2021}. Moreover, a Galactic magnetar has recently been associated with a fast radio burst (FRB) \citep{bochenek2020,chime2020}, which validates the capability of magnetar scenarios to explain at least a fraction of FRBs. 

The pulsed X-ray activity of magnetars shows that they are characterised by a slow rotation period of $\SI{2}{}-\SI{12}{s}$ and a fast spin-down. Under the assumption of a magnetic dipole spin-down, magnetars are therefore constrained to exhibit strong dipolar surface magnetic fields ranging from \SI{e14}{} to \SI{e15}{G} \citep{kouveliotou1999,kaspi2017}, which are two orders of magnitude larger than in regular neutron stars. Furthermore, several lines of evidence suggest the presence of a non-dipolar magnetic field stronger than the dipolar component. Indeed, absorption lines have been detected in the X-ray spectra of two magnetars: SGR 0418+5729 \citep{tiengo2013} and SWIFT J1882.3--1606 \citep{rodriguez2016}. If these lines are interpreted as proton cyclotron lines, they are respectively the signature of non-dipolar magnetic fields of $\sim\SI{2e14}-\SI{e15}{G}$ and $\sim\SI{6e14}-\SI{2.5e15}{G}$, which are stronger than their respective dipolar components by a factor of $\sim 30-170$ \citep{rea2010,rea2012}. Another sign of strong non-dipolar magnetic fields is the detection of a phase modulation in the hard-X-ray emission of a few magnetars. This may be explained by precession movements due to an internal toroidal magnetic field reaching a strength of $\sim\SI{e16}{G}$ \citep{makishima2014,makishima2016,makishima2019,makishima2021}.

Proto-magnetars may be the central engine of extreme events if they are born rotating with a period of a few milliseconds. Indeed, their large-scale magnetic field can extract a large amount of rotational energy, which may create jets and lead to magnetorotational explosions~\citep{ burrows2007,dessart2008,takiwaki2009,kuroda2020,bugli2020,bugli2021,kuroda2020,Obergaulinger2020,Obergaulinger2021,Obergaulinger2022}.
This process may explain hypernovae that are associated with long gamma-ray bursts\citep{duncan1992,zhang2001,woosley2006,drout2011,nomoto2011,gompertz2017,metzger2011,metzger2018}. Moreover, their spin-down luminosity is invoked as a source of delayed energy injection to explain superluminous supernovae (SNe) \citep{woosley2010,kasen2010,dessart2012,inserra2013,nicholl2013}. Finally, millisecond magnetars, which may be formed in binary neutron star mergers, could also provide an explanation for the plateau phase in the X-ray emission of some short gamma-ray bursts \citep{metzger2008,lu2014,gompertz2014}.

\par The central question to understand magnetar formation is the origin of their ultra-strong magnetic field.
One type of scenario invokes magnetic flux conservation during the collapse of magnetised progenitors \citep{ferrario2006,hu2009}. The magnetic field of these progenitors can originate from either a fossil field \citep{braithwaite2004,braithwaite2017} or dynamo action during main sequence star mergers \citep{schneider2019,schneider2020}. While the surface magnetic field is constrained by observations \citep{petit2019}, the magnetic field intensity in the iron core remains unknown, which makes this scenario uncertain. Another class of formation scenarios is the in situ amplification of the magnetic field by a dynamo process after the core collapse, especially at early stages of the proto-magnetar evolution.
Two mechanisms have been studied so far: the convective dynamo \citep{thompson1993,Raynaud2020,raynaud2022,masada2022,white2022} and the magnetorotational instability (MRI)-driven dynamo \citep[e.g.][]{obergaulinger2009,moesta2014,guilet2015,reboul2021a,reboul2021b}. The efficiency of these two dynamo mechanisms in the physical conditions relevant to a proto-neutron star (PNS) is still uncertain, in particular because the regime of very high magnetic Prandtl numbers (i.e. the ratio of viscosity to magnetic diffusivity) has not yet been thoroughly explored \citep{guilet2022,lander2021}. Numerical simulations suggest that the efficiency of both dynamos increases for faster PNS rotation \citep{Raynaud2020,raynaud2022,reboul2021a,reboul2021b}, which makes them good candidates to explain the central engine of extreme explosions. However, it may be more challenging for them to explain magnetar formation in standard SNe, which requires slower initial rotation of the PNS. Indeed, the observed SN remnants associated with Galactic magnetars have an ordinary kinetic explosion energy \citep{vink2006,martin2014,zhou2019}. This suggests that most Galactic magnetars are formed in standard SNe, which is consistent with the fact that extreme explosions represent about 1\% of all SNe whereas magnetars constitute at least 10\% of the whole Galactic young neutron-star population \citep{kouveliotou1994,gill2007,beniamini2019}. Under the assumption that all the rotational energy of the PNS is injected into the kinetic energy of the explosion, the kinetic energy of the proto-magnetar must not exceed the standard kinetic energy of a SN explosion of $\SI{e51}{erg}$, which translates into a constraint on its initial rotation period of $\gtrsim\SI{5}{ms}$ \citep{vink2006}.

All things considered, the aforementioned scenarios require that the progenitor core be either strongly magnetised or fast rotating. It remains uncertain as to whether one of these conditions is met in a sufficient number of progenitors. This article presents our investigation of a new scenario wherein magnetars form from a slowly rotating, weakly magnetised progenitor. We consider the situation in which a newly formed PNS is spun up by the matter initially ejected by the SN explosion that remains gravitationally bound to the compact remnant and eventually falls back onto its surface. As the accretion is asymmetric, recent numerical simulations suggest that the fallback can bring a significant amount of angular momentum to the PNS surface \citep{chan2020,stockinger2020,janka2021}. We investigate the possibility that a magnetar may form due to the dynamo action triggered by the spin up from this fallback accretion. In this scenario, the MRI is expected to be stable because the PNS surface rotates faster than the core. The fallback starts roughly $\sim\SI{5}{}-\SI{10}{s}$ after the core bounce \citep{stockinger2020,janka2021}, which may be too late for the development of a convective dynamo. Instead, we suggest that the magnetic field is amplified by another dynamo mechanism: the so-called Tayler-Spruit dynamo, which is driven by the Tayler instability. This instability feeds off a toroidal field in a stably stratified medium due to the presence of an electric current along the axis of symmetry \citep{tayler1973,pitts1985}.
\citet{spruit2002} proposed a first model of a dynamo driven by the Tayler instability in a differentially rotating stably stratified region. This model has received criticism from several authors \citep[see][]{denissenkov2007,zahn2007}, which has been addressed in the alternative description proposed by \citet{fuller2019}. The Tayler-Spruit dynamo has long been elusive in numerical simulations, but recent numerical simulations provide the first numerical evidence for its existence \citep{petitdemange2022}. This dynamo is usually invoked for magnetic field amplification in the context of stellar interior physics, especially because of its suspected implications for angular momentum transport and the magnetic desert in Ap/Bp stars~\citep[e.g.][]{rudiger2010a,szklarski2013,bonanno2017,guerrero2019,ma2019,bonanno2020,jouve2020}. However, this dynamo process has never been studied in the framework of magnetar formation.

In the following, Sect.~\ref{sec:form} presents the scenario in more detail and the formalism used by our model. We describe our results in Sect.~\ref{sec:res} and discuss them in Sect.~\ref{sec:discu}. Finally, we draw conclusions in Sect.~\ref{sec:conclu}.


\section{Mathematical modelling of the scenario}\label{sec:form}
\begin{figure*}[t]
   \centering
   \includegraphics[width=\textwidth]{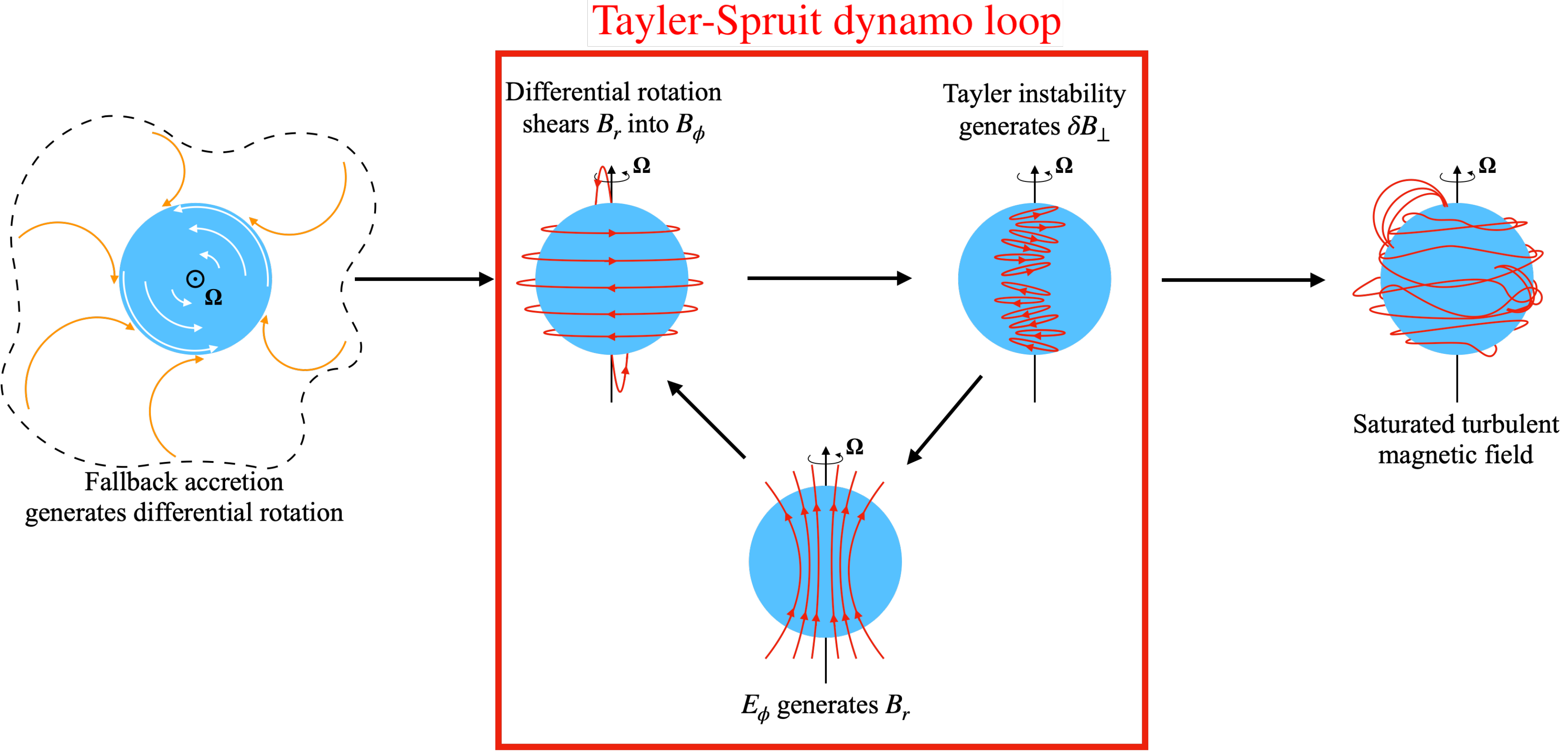}
   \caption{Schematic representation of the different stages of our magnetar formation scenario. The dashed line encloses the region of the fallback (orange arrows). Red and white lines represent the magnetic field lines and fluid motions, respectively. $\vec{\Omega}$ and $E_{\phi}$ stand for the angular rotation frequency and the azimuthal component of the electromotive force, respectively. $B_{\phi}$ and $B_r$ are the axisymmetric azimuthal and radial magnetic field, and $\delta B_{\perp}$ is the non-axisymmetric perpendicular magnetic field.}
              \label{fig:scenario_schematic}%
\end{figure*}

To study our scenario, we built a one-zone model consisting in `average' time evolution equations that capture the main stages sketched in Fig.~\ref{fig:scenario_schematic}. 
We start by describing the impact of the SN fallback on the PNS rotation (the differential rotation) and the magnetic field (the shearing of the radial field and the exponential growth of the Tayler instability). Finally, we present the mathematical formalism for the non-linear stages, that is, the saturation mechanism of the dynamo as modelled by \citet{spruit2002} and \citet{fuller2019}, which we complete by a description of the generation of the radial magnetic field through non-linear induction. For the computation of the time evolution, we only implement the description based on the work of \citet{fuller2019};  but in Sect.~\ref{sec:res} we compare both models regarding the predictions of the saturated magnetic field.

\subsection{Fallback accretion}\label{ssec:sat_reg}
Our scenario starts a few seconds after the core bounce when a fraction of the fallback matter gets accreted onto the PNS surface. This matter is initially ejected during the explosion but stays gravitationally bound to the PNS, and so begins to be asymmetrically accreted \citep{chan2020}. This fallback matter is thought to have a large angular momentum, which can even reach the magnitude of the Keplerian angular momentum \citep{janka2021}. Therefore, the spin of the PNS is strongly affected and the surface rotation can be accelerated up to millisecond periods. In our scenario, the core of the progenitor is assumed to be slowly rotating. Thus, the PNS surface spins faster than the PNS interior, which creates differential rotation.

To model the accretion onto the PNS surface, we use the asymptotic scaling for the mass accretion rate $\dot{M}_{\rm acc} \propto t^{-5/3}$ from \citet{chevalier1989}. As the accretion mass rate must be finite at the beginning of this accretion regime, we define a start time $t_0$ such that
\begin{equation}
    \dot{M}_{\rm acc}=\frac{A}{(t+t_0)^{5/3}}\,,
\end{equation}
where $A$ is a constant. Then, the accreted mass during this regime is 
\begin{equation}
    M_{\rm acc}=\int^{\infty}_0\frac{A}{(t+t_0)^{5/3}}dt\,.
\end{equation}
As $M_{\rm acc}$ is constant, we have
$A=\frac{2}{3}t_0^{2/3}M_{\rm acc}$ and so the accretion mass rate is
\begin{equation} \label{eq:mass_fb}
    \dot{M}_{\rm acc}=\frac{2}{3}M_{\rm acc}\frac{t_0^{2/3}}{(t+t_0)^{5/3}}\,.
\end{equation}

From the fallback matter, only a fraction with angular momentum as large as the Keplerian limit  at most will be accreted by the PNS, as discussed by \citet{janka2021}. Therefore, the relation between the average angular rotation frequency of the PNS and the mass accretion rate is
\begin{equation}\label{eq:omega_dot}
    \dot{\Omega}=\frac{j_{\rm kep}}{I}\dot{M}_{\rm acc}\,,
\end{equation}
where $I$ stands for the PNS moment of inertia and $j_{\rm kep}$~$\equiv$~$\sqrt{GM_\mathrm{PNS}r}$ is the specific Keplerian angular momentum at the PNS surface. As the PNS mass changes little and the contraction of the PNS is almost over at the times considered for the fallback accretion, we assume $I$ to be constant. As supposed in \citet{fuller2019}, the angular momentum is transported faster latitudinally than radially due to stratification, meaning that the differential rotation is shellular, that is $\Omega$ is constant on spherical shells.

As the accretion process spins up only the outer part of the PNS but not its inner core, the shear rate $q\equiv r\partial_r \ln\Omega$ is also expected to evolve. To describe this effect, we use the approximate expression
\begin{equation}\label{shear_evol}
    r\partial_r\Omega \sim \Omega - \Omega(r=0),
\end{equation}
where $\Omega$ and $\Omega(r=0)$ are the average and central angular rotation frequency, respectively. Assuming that the rotation frequency at the centre of the PNS is unchanged by the accretion process (it will change only due to angular momentum transport processes described in Sect.~\ref{ssec:AM}), we infer the time derivative of the shear rate as
\begin{equation}
    \dot{q}\sim\frac{\dot{\Omega}}{\Omega}(1-q).
\end{equation}

\subsection{Shearing and Tayler instability growth} \label{ssec:shear}
The differential rotation generated by the fallback will shear the radial component of the large-scale radial magnetic field $B_r$ into the azimuthal field $B_{\phi}$ as follows
\begin{equation}\label{eq:Bphi_dot}
    \partial_t B_{\phi}=q\Omega B_r\,.
\end{equation}Therefore, we can define a growth rate\footnote{ We note that this growth rate and the others defined below are rather {instantaneous} growth rates because they depend on the magnetic field.} for $B_{\phi}$:
\begin{equation}\label{eq:shear}
    \sigma_{\rm shear}\equiv q\Omega \frac{B_r}{B_{\phi}}\,.
\end{equation}

As $B_{\phi}$ grows, it becomes Tayler unstable. To depict the linear growth of the instability, we make the following assumptions. First, the stratified medium of the PNS interior is characterised by the Brunt-V\"{a}is\"{a}l\"{a} frequency \citep{hudepohl2014}:
\begin{equation}\label{eq:brunt-vaisala}
    N\equiv \sqrt{-\frac{g}{\rho}\left(\left.\frac{\partial\rho}{\partial S}\right|_{P,Y_e}\frac{dS}{dr}+\left.\frac{\partial \rho}{\partial Y_e}\right|_{P,S}\frac{dY_e}{dr}\right)}
    \sim\SI{4e3}{s^{-1}}
    \,,
\end{equation}
where $g$, $\rho$, $Y_e$, and $S$ are the gravitational acceleration, the PNS mean density, the electron fraction, and the entropy, respectively.
In the remainder of this paper, we use the fiducial value $N = \SI{4e3}{s^{-1}}$ based on the results of the 1D core-collapse supernova (CCSN) simulations from \citet[Chap.~5]{hudepohl2014}. 
The Brunt-V\"{a}is\"{a}l\"{a} frequency is almost uniform in most of the PNS except 
near the surface where it peaks at $\sim\SI{e4}{s^{-1}}$. \citet{hudepohl2014} made a comparison between two different equations of state (EOS): Shen \citep{shen1998b,shen1998a,shen2011} and LS220 \citep{lattimer1991}, and found that the choice of the EOS mainly affects the localisation and duration of the convection but not the value of the Brunt-V\"{a}is\"{a}l\"{a} frequency in the stably stratified region.
Second, the main background azimuthal field is $B_{\phi}$, which is associated with the Alfvén frequency: 
\begin{equation}
    \omega_{\rm A}\equiv\frac{B_{\phi}}{\sqrt{4\pi\rho r^2}}\simeq\SI{11.6}{}\left(\frac{B_{\phi}}{\SI{e15}{G}}\right) \SI{}{s^{-1}}
    \,,
\end{equation}
for $r=\SI{12}{km}$ and $\rho=\SI{4.1e14}{g.cm^{-3}}$.
Finally, the frequencies characterising the PNS are ordered such that
\begin{equation}\label{eq:key_freq}
    N\gg\Omega\gg\omega_{\rm A}\,.
\end{equation}
The development of the Tayler instability is triggered after reaching the critical strength~\citep{spruit1999,spruit2002,zahn2007}
\begin{equation}\label{eq:Bcrit_tayler}
\begin{split}
    B_{\phi}>B_{\phi\rm, c} & \sim\Omega\left(\frac{N}{\Omega}\right)^{1/2}\left(\frac{\eta}{r^2\Omega}\right)^{1/4}\sqrt{4\pi\rho r^2} \\
    & \simeq\SI{2.5e12}{}\left(\frac{\Omega}{\SI{200\pi}{rad.s^{-1}}}\right)^{1/4}\SI{}{G}\,,
\end{split}
\end{equation}
where $\eta\sim\SI{e-4}{cm^2.s^{-1}}$ \citep{thompson1993} is the magnetic diffusivity. The fastest-growing perturbations are the $m=1$ modes with an associated rate of \citep{ma2019}
\begin{equation}\label{eq:growth_TI}
    \sigma_{\rm TI}\sim\frac{\omega_{\rm A}^2}{\Omega}\simeq\SI{0.21}{} \left(\frac{B_{\phi}}{\SI{e15}{G}}\right)^2 \left(\frac{\Omega}{\SI{200\pi}{rad.s^{-1}}}\right)^{-1} \SI{}{s^{-1}}\,.
\end{equation}
As the PNS interior is strongly stratified, we can determine a maximum radial length scale for the instability
\begin{equation}\label{eq:strat}
    l_r\sim\frac{\omega_{\rm A}}{N}l_{\perp}\,,
\end{equation}
where the horizontal length scale is approximated by $l_{\perp}\sim r$.

\subsection{Spruit's picture of the dynamo} \label{ssec:spruit}

\citet{spruit2002} proposes that the energy in the azimuthal large-scale field $B_{\phi}$ cascades to small scale, that is the form of the non-linear magnetic energy dissipation is
\begin{equation}\label{eq:spruit_damp}
    \dot{E}_{\rm mag}\sim\gamma_{\rm turb}|B_{\phi}|^2\,,
\end{equation}
where $\gamma_{\rm turb}$ is the turbulent damping rate. To determine this rate, \citet{spruit2002} argues that the saturation of the instability occurs when the turbulent velocity field generates a sufficiently large effective turbulent diffusivity to balance the growth rate of the instability, that is
\begin{equation}\label{eq:spruit_damp_rate}
    \gamma_{\rm turb}\sim\frac{\eta_{\rm e}}{l_r^2}\sim\sigma_{\rm TI}\,,
\end{equation}
where $\eta_{\rm e}$ is an effective turbulent diffusivity. 

The solenoidal character of the perturbed magnetic field implies $B_r/l_r\sim B_{\phi}/l_{\perp}$, which leads to
\begin{equation}\label{eq:spruit_indu}
    B_r\sim\frac{\omega_{\rm A}}{N}B_{\phi}\,,
\end{equation}
using the relation between length scales of the instability given by Eq.~\eqref{eq:strat}.
As the azimuthal magnetic field $B_{\phi}$ is generated via the shear of the radial magnetic field $B_r$, the dynamo is expected to saturate when the shear (Eq.~\ref{eq:shear}) balances the turbulent damping (Eq.~\ref{eq:spruit_damp_rate}). Thus, the amplitudes of the magnetic field components saturate at
\begin{align}
    B_{\phi,\rm S}^{\rm sat}&\sim \sqrt{4\pi\rho r^2}q\frac{\Omega^2}{N}\,,\label{eq:Bphi_spruit}\\ 
    B_{r,\rm S}^{\rm sat}&\sim \sqrt{4\pi\rho r^2}q^2\frac{\Omega^4}{N^3}\,. 
    \label{eq:Br_spruit}
\end{align}
This description of the dynamo mechanism has been criticised for two reasons:   First of all, if the large-scale component of $B_{\phi}$ remains constant on larger length scales than $l_r$, the displacements produced by the instability are not expected to mix the large-scale field lines to damp $B_{\phi}$ through reconnection. Therefore, the damping rate estimated in Eq.~\eqref{eq:spruit_damp_rate} is overestimated for the large-scale components of the azimuthal field $B_{\phi}$ \citep[see][]{fuller2019}. Secondly, as $m=1$ modes are dominant, the radial magnetic field $B_r$ produced by the instability is non-axisymmetric, and therefore its shear generates a mostly non-axisymmetric azimuthal field $B_{\phi}$.
    Hence, the axisymmetric component of the fields $B_r$ and $B_{\phi}$ may not be related by Eq.~\eqref{eq:spruit_indu} \citep[see][]{zahn2007}.

\subsection{A revised model of the dynamo} \label{ssec:fuller}
This section presents a description of the dynamo that completes the model proposed by \citet{fuller2019} in the sense that we consider the time evolution of the magnetic field. A clear distinction is now made between the `axisymmetric' components $B_{\phi}$, $B_r$, and the `non-axisymmetric' perturbed components $\delta B_{\perp}$, $\delta B_r$, which become the ones connected by the solenoidal condition
\begin{equation}\label{eq:soleno}
    \delta B_r\sim\frac{\omega_{\rm A}}{N}\delta B_{\perp}\,.
\end{equation}
To overcome the previously raised difficulties, \citet{fuller2019} argue that the energy in the perturbed field $\vec{\delta B}$ dissipates to small scales and find that the damping rate is
\begin{equation}\label{eq:cas}
    \gamma_{\rm cas}\sim\frac{\delta v_{\rm A}}{r}\,,
\end{equation}
where $\delta v_{\rm A}\equiv\delta B_{\perp}/\sqrt{4\pi\rho}$ is the perturbed Alfvén velocity. Thus, equating the instability growth rate (Eq.~\ref{eq:growth_TI}) and the damping rate (Eq.~\ref{eq:cas}) gives the saturation strength of the perturbed field $\delta B_{\perp}$ for a given strength of azimuthal field $B_{\phi}$,
\begin{equation}\label{eq:sat_TI}
    \delta B_{\perp}\sim\frac{\omega_{\rm A}}{\Omega}B_{\phi}\,.
\end{equation}
When the instability is saturated, the non-linear magnetic energy dissipation is then
\begin{equation}\label{eq:emag1}
     \dot{E}_{\rm mag}\sim\gamma_{\rm cas}|\vec{\delta B}|^2\sim\frac{\delta v_{\rm A}}{r}\left|{\delta B_{\perp}}\right|^2\,.
\end{equation}
As the azimuthal field $B_{\phi}$ is the dominant magnetic component, $\dot{E}_{\rm mag}\sim B_{\phi}\partial_tB_{\phi}$. Hence, a damping rate can be defined for the axisymmetric components $B_{\phi}$ and $B_{r}$:
\begin{equation}\label{eq:diss1}
    \gamma_{\rm diss}\equiv\frac{\dot{E}_{\rm mag}}{B_{\phi}^2}\,.
\end{equation}
As the previous expression of the magnetic energy (Eq.~\ref{eq:emag1}) is only valid when the instability saturates, we use the expression
\begin{equation}\label{eq:emag2}
    \dot{E}_{\rm mag}\sim \frac{\omega_{\rm A}^2}{\Omega}|\delta B_{\perp}|^2\,,
\end{equation}
which is valid in both the saturated and non-saturated states. Therefore, the damping rate defined in Eq. \eqref{eq:diss1} becomes
\begin{equation}\label{eq:diss2}
     \gamma_{\rm diss}\sim\frac{\omega_{\rm A}^2}{\Omega}\left(\frac{\delta B_{\perp}}{B_{\phi}}\right)^2\,.
\end{equation}
To close the dynamo loop, the Tayler instability must generate the axisymmetric radial magnetic field $B_r$ ($\alpha$-effect), which will be sheared again ($\Omega$-effect). In the framework of the mean field theory, the induction equation reads
\begin{equation}\label{eq:indu}
    \partial_t\langle\vec{B}\rangle = \mathbf{\nabla}\times\left(\langle\vec{v}\rangle\times\langle\vec{B}\rangle + \vec{E  } \right)
    - \eta\Delta\langle\vec{B}\rangle\,,
\end{equation}
in which we ignore the resistive term. Considering the average symbol $\langle\mathbf{\cdot}\rangle$ as an azimuthal average, we note $\langle\vec{B}\rangle=\vec{B}$ in order to remain consistent with our notation of the axisymmetric magnetic field.
The electromotive force $\vec{E}\equiv\left\langle\delta\vec{v}\wedge\delta\vec{B}\right\rangle$ is the important non-linear quantity responsible for the generation of the axisymmetric radial field $B_r$. In spherical coordinates, the radial component of Eq.~\eqref{eq:indu} is
\begin{equation}\label{eq:indu_1}
    \partial_t B_r=\frac{1}{r \sin\theta}\left[\partial_{\theta}(\sin\theta\, E_{\phi})-\partial_{\phi}E_{\theta}\right]\,.
\end{equation}
As $B_r$ is axisymmetric, $E_{\theta}$ can be ignored. By definition, the azimuthal component of electromotive force is
\begin{equation}\label{eq:EMFp}
    E_{\phi}=\delta v_r\delta B_{\theta}-\delta v_{\theta}\delta B_r\,.
\end{equation}
Supposing an incompressible perturbed velocity field and using Eq.~\eqref{eq:soleno}, we write
\begin{equation}\label{eq:incomp}
    \frac{\delta v_r}{\delta v_{\perp}}\sim\frac{\delta B_r}{\delta B_{\perp}}\sim\frac{\omega_{\rm A}}{N}
    \,,
\end{equation}
and so the azimuthal electromotive force reads
\begin{equation}
    E_{\phi}\sim\delta v_{\theta}\delta B_{r}\sim\delta v_r\delta B_{\theta}\sim\delta v_r\delta B_{\perp}\,,
\end{equation}
where we assume that the two terms on the right-hand side of Eq.~\eqref{eq:EMFp} do not cancel. The production of the radial field $B_r$ can be approximated by
\begin{equation}
    \partial_tB_r\sim \frac{E_{\phi}}{r}\sim\frac{\delta v_r\delta B_{\perp}}{r}\,.
\end{equation}
We must note that this expression differs from Eq.$\,$(29) in \citet{fuller2019}, which appears to contain a typo. As in \citet{fuller2019} and \citet{ma2019}, we expect magnetostrophic balance $\delta v_{\perp}\sim\delta v_{\rm A}\omega_{\rm A}/\Omega$, which leads to
\begin{equation}
    \delta v_r\sim\frac{\omega_{\rm A}^2}{N\Omega}\delta v_{\rm A}\,.
\end{equation}
Thus,
\begin{equation}
    \partial_t B_r\sim\frac{E_{\phi}}{r}\sim\frac{\omega_{\rm A}^2}{N\Omega}\frac{\delta B_{\perp}^2}{\sqrt{4\pi\rho r^2}}\,,
\end{equation}
and we can define a growth rate for $B_r$
\begin{equation}\label{eq:induc}
    \sigma_{\rm NL}\equiv \frac{1}{\sqrt{4\pi\rho r^2}}\frac{\omega_{\rm A}^2}{N\Omega}\frac{\delta B_{\perp}^2}{B_r}\,.
\end{equation}
The radial field $B_r$ will saturate when its non-linear growth rate (Eq.~\ref{eq:induc}) is balanced by the turbulent dissipation (Eq.~\ref{eq:diss2}). This way, we find the relation between the axisymmetric fields
\begin{equation}\label{eq:sat_dyn}
    B_r\sim\frac{\omega_{\rm A}}{N}B_{\phi}\,,
\end{equation}
using Eq.~\eqref{eq:sat_TI}.
We note that this relation is similar to Eq.~\eqref{eq:spruit_indu} from \citet{spruit2002}, which was derived for the non-axisymmetric components only. \citet{fuller2019} also established the same relation arguing that the Tayler instability cannot operate when the magnetic tension forces become larger than the magnetic pressure forces leading to the instability.

The azimuthal magnetic field saturates when the shear rate (Eq.~\ref{eq:shear}) balances the dissipation rate (Eq.~\ref{eq:diss2}). Thus, using the relations between the magnetic field components (Eqs.~\ref{eq:sat_TI} and \ref{eq:sat_dyn}), we find the magnetic field strengths in the saturated regime derived in \citet{fuller2019}:
\begin{align}
    B_{\phi,\rm F}^{\rm sat}&\sim\sqrt{4\pi\rho r^2}\Omega\left(\frac{q\Omega}{N}\right)^{1/3}\,, \label{eq:Bphi_sat}\\
    \delta B^{\rm sat}_{\perp,\rm F}&\sim\sqrt{4\pi\rho r^2}\Omega\left(\frac{q\Omega}{N}\right)^{2/3}\,, \label{eq:Bperp_sat}\\
    B^{\rm sat}_{r,\rm F}&\sim\sqrt{4\pi\rho r^2}\Omega\left(\frac{q^2\Omega^5}{N^5}\right)^{1/3}\,.\label{eq:Br_sat}
\end{align}

Finally, the angular momentum is redistributed in the PNS through Maxwell stresses associated with an effective angular momentum diffusivity \citep{spruit2002,fuller2019}:
\begin{equation}\label{eq:nu_AM}
    \nu_{\rm AM}=\frac{B_rB_{\phi}}{4\pi\rho q \Omega}\,,
\end{equation}
which affects the shear parameter at the rate
\begin{equation}\label{eq:AM_trans}
    \gamma_{\rm AM}\equiv\frac{\nu_{\rm AM}}{r^2}\,.
\end{equation}


\subsection{Governing evolution equations}\label{ssec:gov_eq}
Now that the main equations involved in our scenario have been brought out, we can write the evolution equations for the rotation properties and the magnetic field. The evolution of PNS angular rotation frequency is driven by the fallback accretion rate (Eq.~\ref{eq:mass_fb}) as described by Eq.~\eqref{eq:omega_dot}. Hence,
\begin{equation}\label{eq:omg_evol}
    \dot{\Omega}=\frac{2}{3}\Delta\Omega\frac{t_0^{2/3}}{(t+t_0)^{5/3}}\,,
\end{equation}
where $\Delta \Omega = {\Omega}_{\rm fin}-{\Omega}_{\rm init} = M_{\rm acc}j_{\rm kep}/I$. As previously mentioned, the shear rate is also expected to decrease due to angular momentum transport (Eq.~\ref{eq:AM_trans}) such that
\begin{equation}\label{eq:q_evol}
    \dot{q}=\frac{\dot{\Omega}}{\Omega}(1-q)-\gamma_{\rm AM}q=\frac{2}{3}\frac{\Delta\Omega}{\Omega}\frac{t_0^{2/3}}{(t+t_0)^{5/3}}-\frac{B_rB_{\phi}}{4\pi\rho\Omega r^2}\,.
\end{equation}
Combining the different growth and damping rates given by Eqs.~\eqref{eq:shear}, \eqref{eq:growth_TI}, \eqref{eq:cas}, \eqref{eq:diss2}, and \eqref{eq:induc}, we find that the magnetic field evolution is governed by the following equations:
\begin{align}
    \partial_t B_{\phi}&=\left(\sigma_{\rm shear}-\gamma_{\rm diss}\right)B_{\phi}=q\Omega B_r-\frac{\omega_{\rm A}^2}{\Omega}\frac{\delta B_{\perp}^2}{B_{\phi}}\,,\label{eq:Bphi_evol}\\
    \partial_t\delta B_{\perp}&=\left(\sigma_{\rm TI}-\gamma_{\rm cas}\right)\delta B_{\perp}=\frac{\omega_{\rm A}^2}{\Omega}\delta B_{\perp}-\frac{\delta v_{\rm A}}{r}\delta B_{\perp}\,,\label{eq:Bperp_evol}\\
    \partial_t B_r&=\left(\sigma_{\rm NL}-\gamma_{\rm diss}\right)B_{r}=\frac{\omega_{\rm A}^2}{N\Omega}\frac{\delta B_{\perp}^2}{\sqrt{4\pi\rho r^2}}-\frac{\omega_{\rm A}^2}{\Omega}\left(\frac{\delta B_{\perp}}{B_{\phi}}\right)^2B_r\,.\label{eq:Br_evol}
\end{align}

Equations~\eqref{eq:omg_evol}--\eqref{eq:Br_evol} are solved for a typical PNS of 
\SI{5}{}-\SI{10}{s} in age using the odeint function from the Python package SciPy. The PNS mass and radius are $M_{\rm PNS}=\SI{1.5}{M_{\odot}}$ and $r=\SI{12}{km}$, and so the average density is $\rho=\SI{4.1e14}{g.cm^{-3}}$. The moment of inertia is estimated using Eq.~(12) from \citet{lattimer2005}:
\begin{equation}
\begin{split}
    I & =0.237M_{\rm PNS}r^2\left[1+4.2\left(\frac{M_{\rm PNS}}{\rm M_{\odot}}\frac{\SI{1}{km}}{r}\right)\right.
    \left.+90\left(\frac{M_{\rm PNS}}{\rm M_{\odot}}\frac{\SI{1}{km}}{r}\right)^4\right] \\ &\simeq\SI{1.6e45}{g.cm^2}\,.
\end{split}
\end{equation}

The PNS core is assumed to be initially in solid-body rotation (i.e. $q=0$) and slowly rotating with an angular rotation frequency $\Omega_{\rm init}=\SI{2\pi}{rad.s^{-1}}$ (i.e. an initial rotation period  $P_{\rm init}\equiv 2\pi/\Omega_{\rm init}=\SI{1}{s}$). We note that the results of the time integration weakly depend on these parameters as long as $P_{\rm init}\gtrsim \SI{40}{ms}$. The parameters of the fallback are chosen to be consistent with the simulations of \citet{janka2021}, with a starting time at $t_0=\SI{7}{s}$. The initial magnetic field components $B_r$, $B_\phi$, and $\delta B_\perp$ are initialised at a strength of \SI{e12}{G}, which is the typical magnetic field amplitude of regular neutron stars.

\section{Results}\label{sec:res}
We proceed with a twofold presentation of our model outputs: First we present the time evolution, in which we derive analytical predictions for the timescales of the different phases of the scenario and compare them to the integrated evolution. We then present the saturated regime where we focus on the magnetic field intensities reached via the Tayler-Spruit dynamo and provide an `upper' limit of PNS rotation period (i.e. a `lower' limit of fallback mass) to form magnetars.

\subsection{Time evolution of the magnetic field}\label{ssec:time_evolution}
\begin{figure}[tbp]
   \centering
   \includegraphics[width=0.49\textwidth]{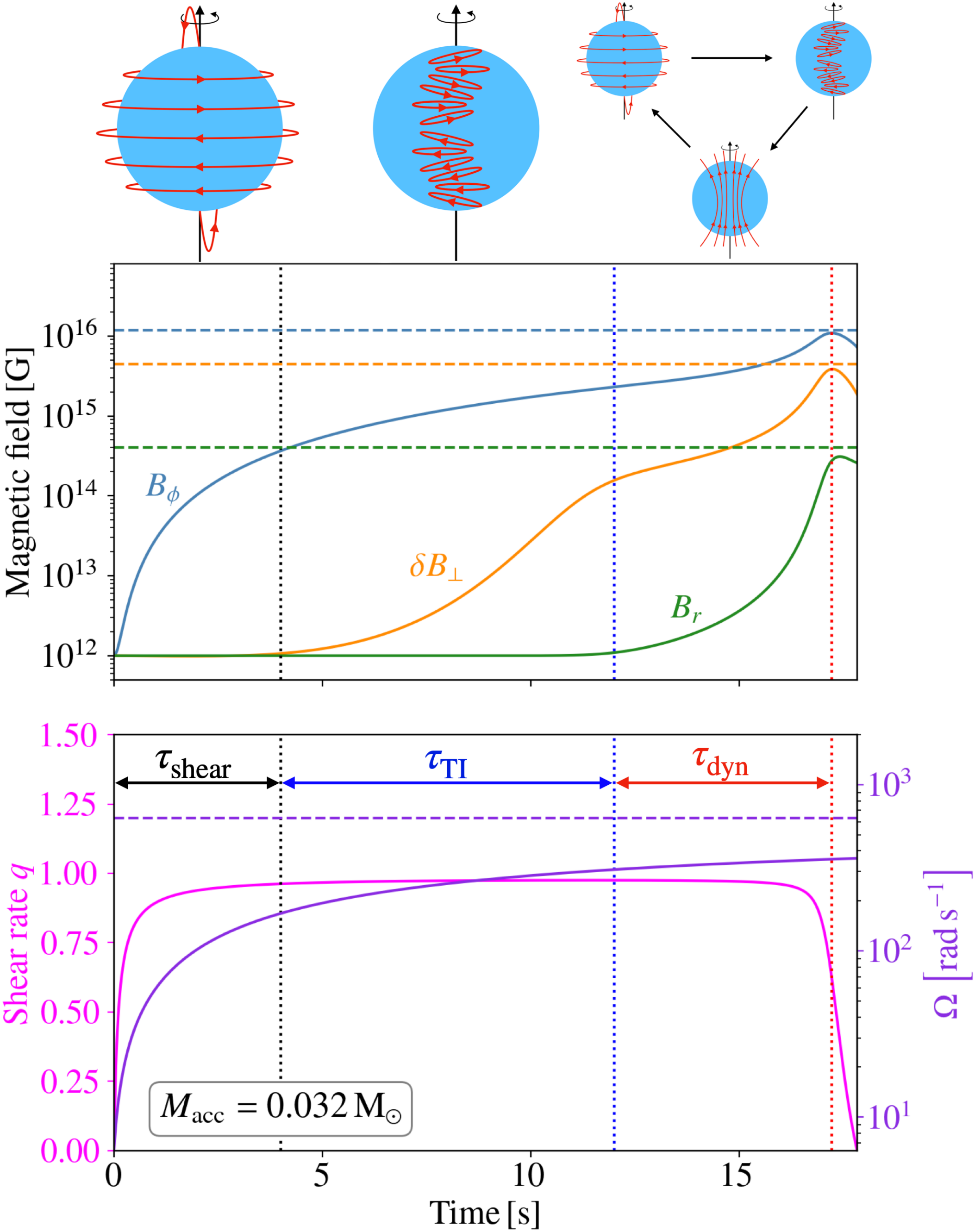}
   \caption{Time evolution of the different components of the magnetic field (top), the dimensionless shear rate, and the angular rotation frequency (bottom) for an accreted mass $M_{\rm acc}=\SI{3.2e-2}{M_{\odot}}$ corresponding to an asymptotic rotation period $P_{\rm fin}=\SI{10}{ms}$. The different stages of the dynamo process are highlighted by the schematics at the top and their associated timescales by the double arrows. Their ends are illustrated by the dotted vertical lines: winding (black), linear development of the Tayler instability (dark blue), and the whole dynamo loop (red). The horizontal dashed lines (blue, orange, and green) show respectively the saturation intensities $B^{\rm sat}_{\phi,\rm F}, \delta B^{\rm sat}_{\perp,\rm F}$, and $B^{\rm sat}_{r,\rm F}$ (Eqs.~\ref{eq:Bphi_sat}--\ref{eq:Br_sat}) evaluated at the time of saturation. The  dashed violet horizontal line represents the asymptotic angular rotation frequency $\Omega_{\rm fin}$.}\label{fig:typical_case}%
\end{figure}

The time series for an asymptotic rotation period $P_{\rm fin}\equiv 2\pi/\Omega_{\rm fin}=\SI{10}{ms}$ displayed in Fig.~\ref{fig:typical_case} can be split into several phases, which are illustrated by the schematics at the top of the figure: 
\begin{itemize}
    \item[(i)] $B_{\phi}$ is strongly amplified for $\sim \SI{4}{s}$ due to the winding of $B_r$ while the other components stay constant. As the mass-accretion rate is higher in this phase, strong increases in the shear rate and the rotation rate are also noted.
    \item[(ii)] The Tayler instability develops and amplifies $\delta B_{\perp}$ for $\sim$~$\SI{8}{s}$.
    \item[(iii)] The axisymmetric radial field $B_r$ is generated allowing the dynamo action to occur for $\sim$~$\SI{5}{s}$. The azimuthal magnetic field saturates at $\sim\SI{17}{s}$, which corresponds to $\sim$~$\SI{24}{s}$ after the core bounce.
    \item[(iv)] The angular momentum transport by the magnetic field, which was negligible during the first $\sim$~$\SI{16}{s}$, becomes significant as the magnetic field saturates. This stage is discussed in Sect.~\ref{ssec:AM}.
\end{itemize}
The evolution of the magnetic field is shown until the whole angular momentum is transported, (i.e. when the shear rate reaches $q=0$) at $t\sim\SI{17.5}{s}$. Further evolution is not considered because our set of equations does not intend to describe either the relaxation phase of the magnetic field to a stable geometric configuration or the dynamics with very low shear where one would expect the Tayler-Spruit dynamo to stop or to act intermittently \citep{fuller2022}.

\begin{figure*}[tbp]
    \centering
    \includegraphics[width=0.45\textwidth]{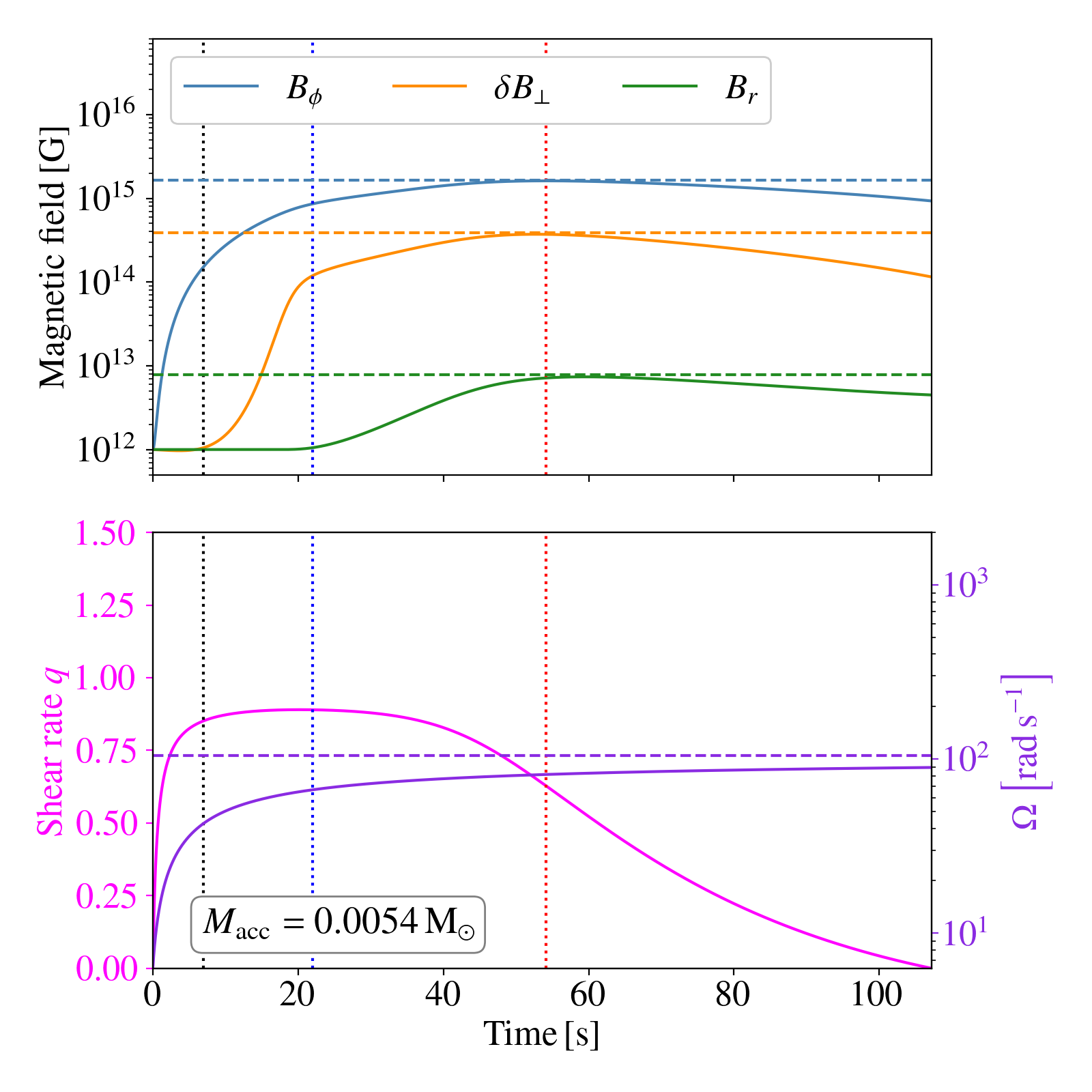}
    \includegraphics[width=0.45\textwidth]{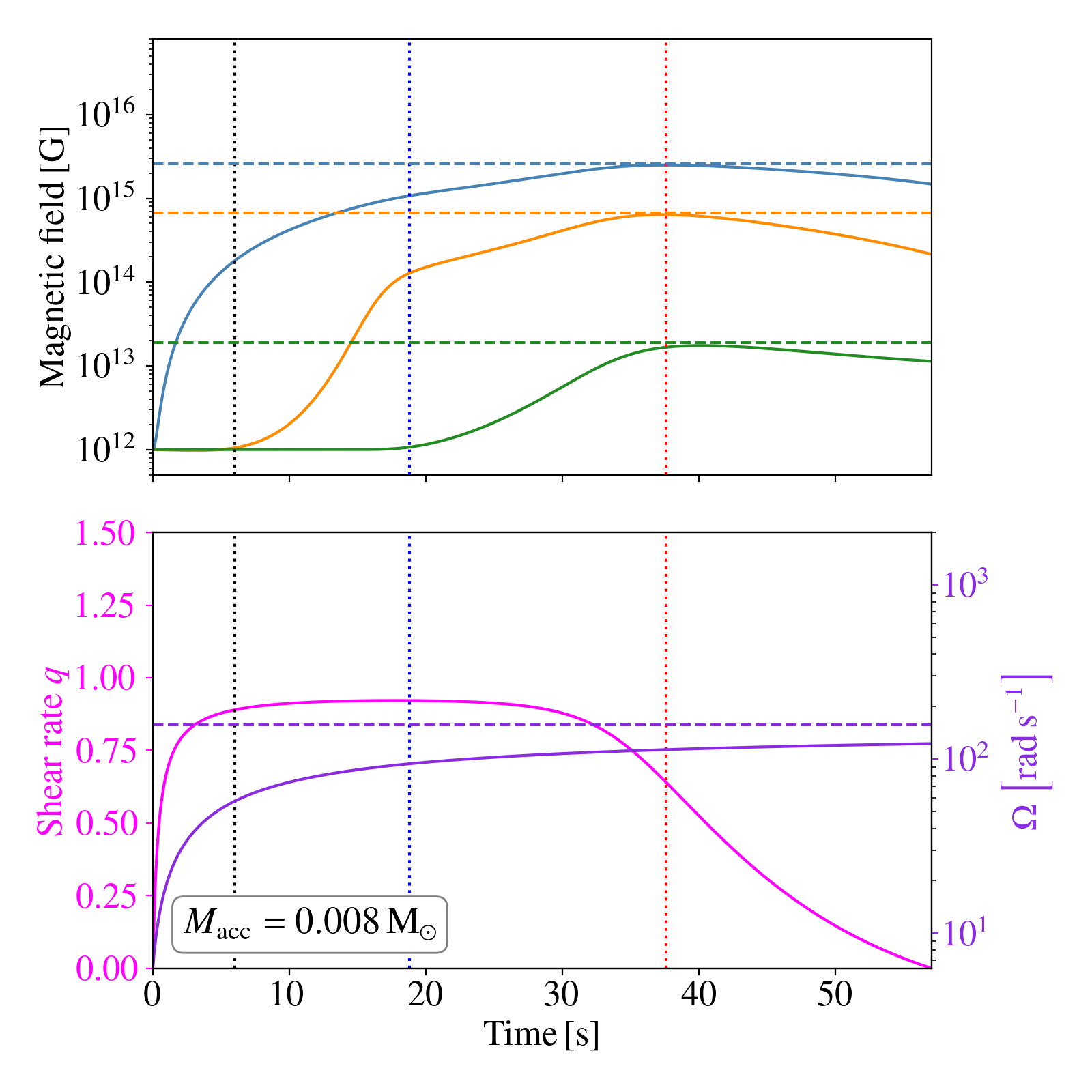}
    \includegraphics[width=0.45\textwidth]{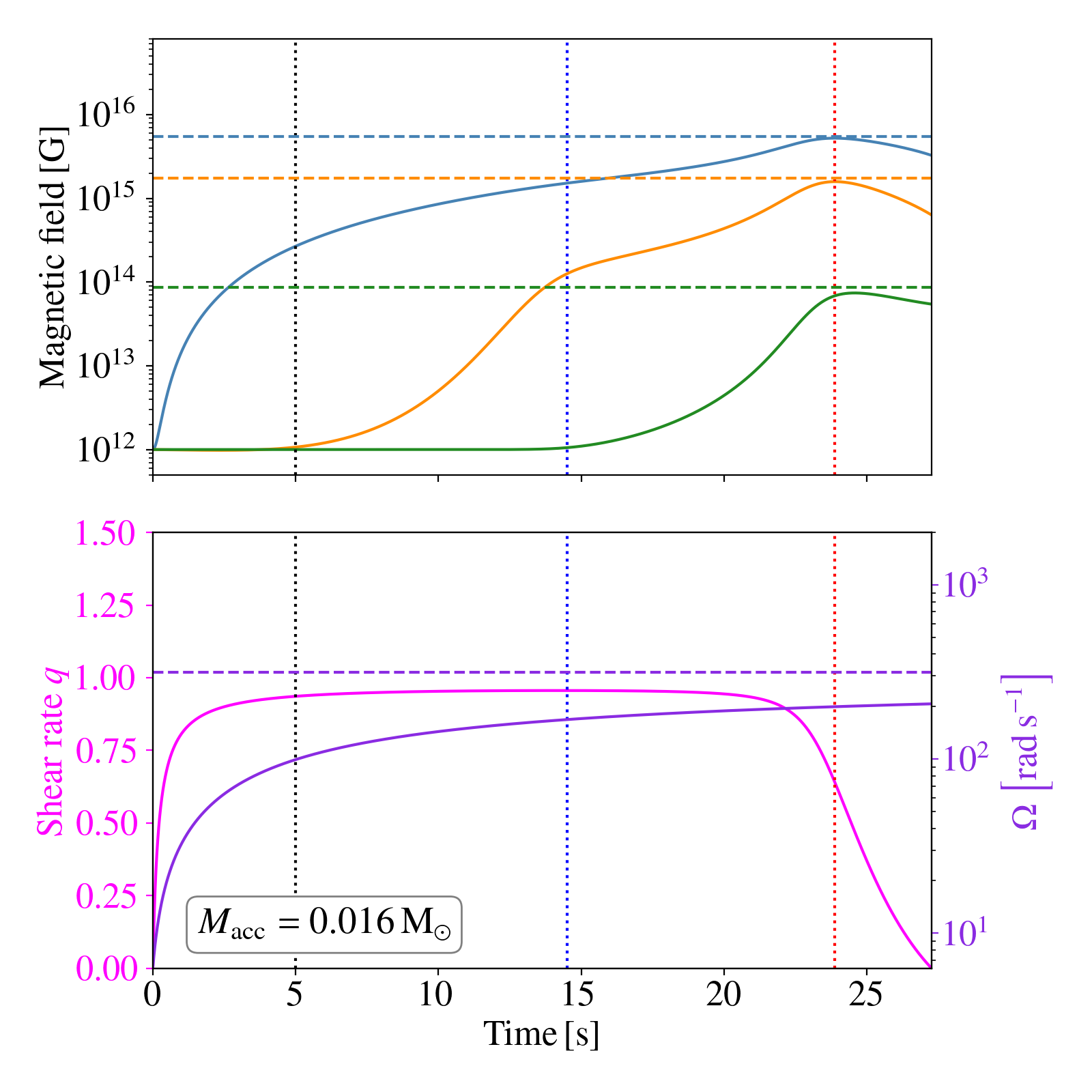}
    \includegraphics[width=0.45\textwidth]{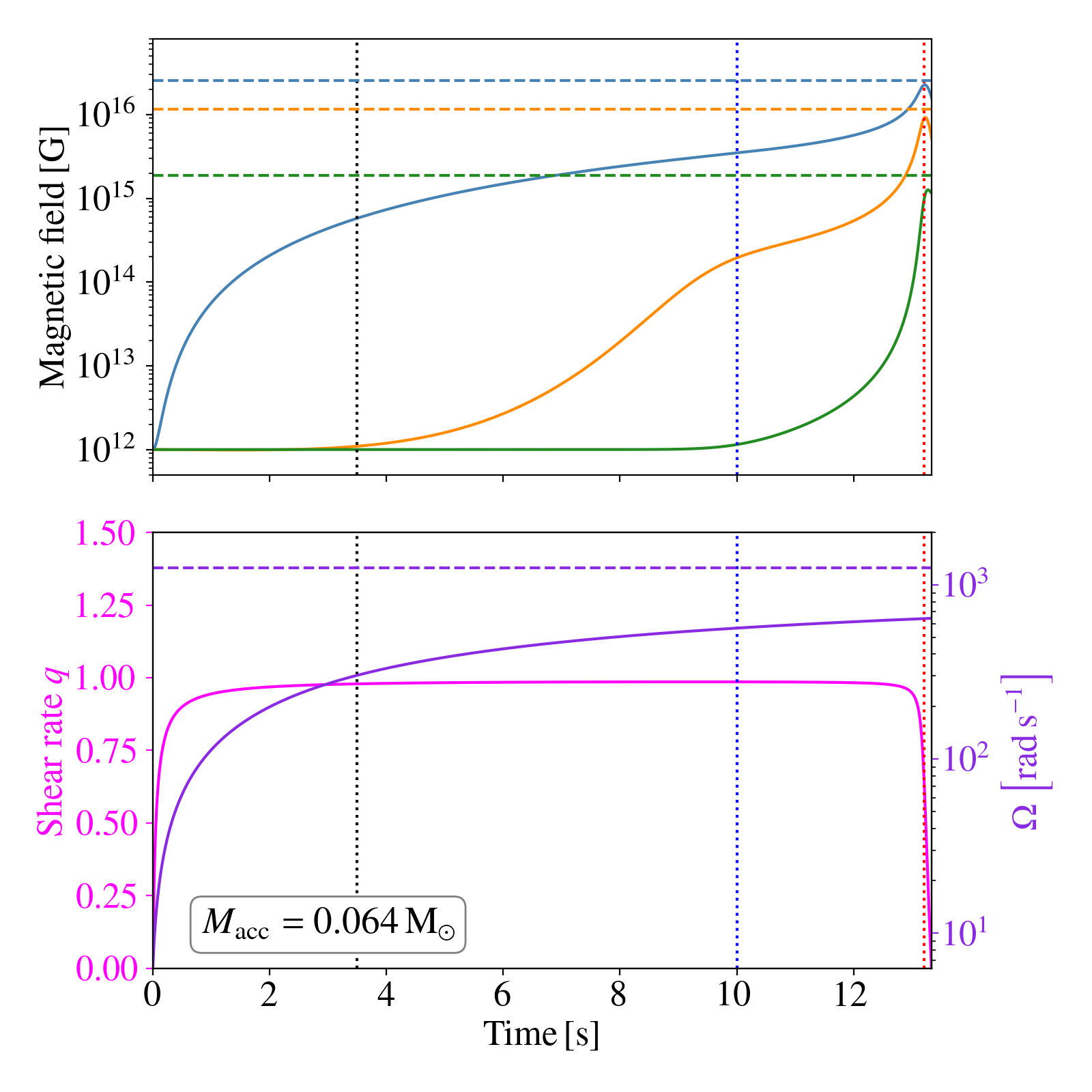}
    \caption{Same as Fig.~\ref{fig:typical_case} but for total accreted masses of $M_{\rm acc}=\{\SI{0.54}{}$, $\SI{0.8}{}$, $\SI{1.6}{}$, $\SI{6.4}{}\}\times\SI{e-2}{{\rm M}_{\odot}}$ (corresponding to $P_{\rm fin}=\{\SI{60}{}$, $\SI{40}{}$, $\SI{20}{}$, $\SI{5}{}\}\SI{}{ms}$, respectively).}
    \label{fig:diff_mass}
\end{figure*}

The different vertical lines in Fig.~\ref{fig:diff_mass} show that the above phases occur at different times for different accreted masses. To better quantify the dependence on this parameter, we derive analytical estimates of the corresponding characteristic timescales $\tau_{\rm shear}$, $\tau_{\rm TI}$, and $\tau_{\rm dyn}$ (see Fig.~\ref{fig:typical_case}). First, the shearing phase begins when the fallback matter starts to be accreted on the PNS surface and finishes when the azimuthal magnetic field $B_{\phi}$ is strong enough to make the Tayler instability grow as fast as $B_{\phi}$, that is when the growth rate of the instability (Eq.~\ref{eq:growth_TI}) is equal to the winding rate (Eq.~\ref{eq:shear}). Thus, the Alfvén frequency associated with the intensity of the azimuthal magnetic field at the end of this phase is
\begin{equation}\label{eq:omega_TI}
    \omega_{\rm A}\sim \omega_{\rm A, TI}\equiv \left(q\Omega^2\omega_{r,0}\right)^{1/3}\,,
\end{equation}
where $\omega_{r,0}\equiv B_r(t=0)/\sqrt{4\pi\rho r^2}$. Therefore, a characteristic timescale for the shearing stage can be defined as the inverse of the winding growth rate (Eq.~\ref{eq:growth_TI}) evaluated at $\omega_{\rm A}=\omega_{\rm A, TI}$:
\begin{equation}\label{eq:t_shear}
    \tau_{\rm shear}\equiv\sigma_{\rm shear}^{-1} \sim\frac{\omega_{\rm A,\rm TI}}{q\Omega \omega_{r,0}}=\left(q^2\omega_{r,0}^2\Omega\right)^{-1/3}\,.
\end{equation}

Second, as the azimuthal field becomes unstable, the Tayler instability grows exponentially until the perturbed field reaches the saturation intensity of Eq.~\eqref{eq:sat_TI}. The perturbed field at saturation can be approximated by 
\begin{equation}
    \delta B_{\perp}(t=t_{\rm sat})\sim\delta B_{\perp}(t=\tau_{\rm shear})\,{\rm exp}\left[\sigma_{\rm TI}(t_{\rm sat}-\tau_{\rm shear})\right]\,, 
\end{equation}
and so a characteristic timescale for this stage can be defined as
\begin{equation}
    \tau_{\rm TI}\equiv t_{\rm sat}-\tau_{\rm shear}\sim\sigma_{\rm TI}^{-1}\,{\rm ln}\left[\frac{\delta B_{\perp}(t=t_{\rm sat})}{\delta B_{\perp}(t=\tau_{\rm shear})}\right]\,.
\end{equation}
Using Eq.~\eqref{eq:sat_TI}, we have
\begin{equation}
    \delta B_{\perp}(t=t_{\rm sat})\sim \frac{\omega_{\rm A}(t=t_{\rm sat})}{\Omega}B_{\phi}(t=t_{\rm sat})\,.
\end{equation}
In order to obtain a simple estimate, we make the rough approximations that $\delta B_{\perp}(t=\tau_{\rm shear})\sim\delta B_{\perp}(t=0)$ and $B_{\phi}(t=\tau_{\rm shear})\sim B_{\phi}(t=t_{\rm sat})$, which leads to
\begin{equation}\label{eq:t_TI}
    \tau_{\rm TI}\sim\frac{\Omega}{\omega_{\rm A, TI}^2} {\rm ln}\left(\frac{\omega_{\rm A, TI}^2r}{\Omega\delta v_{\rm A,0} }\right)\,,
\end{equation}
where $\delta v_{\rm A,0}\equiv\delta v_{\rm A}(t=0)$. 

Third, when the perturbed field reaches a sufficient amplitude, the axisymmetric radial field is amplified through non-linear induction, thus closing the dynamo loop. This phase ends when the magnetic field
saturates at the intensities given by Eqs.~\eqref{eq:Bphi_sat}--\eqref{eq:Br_sat}. Likewise, we estimate the critical strength of the azimuthal field $B_{\phi,\rm dyn}$ above which the dynamo loop is triggered by equating the growth rate of the radial field $B_{r}$ (Eq.~\ref{eq:induc}) and the winding rate (Eq.~\ref{eq:shear}). We obtain the Alfvén frequency associated with $B_{\phi,\rm dyn}$
\begin{equation}
    \omega_{\rm A, dyn}\equiv\left(qN\Omega^4\omega_{r,0}^2\right)^{1/7}\,,
\end{equation}
making use of Eq.~\eqref{eq:sat_TI}. We define the dynamo characteristic timescale as\begin{equation}
    \tau_{\rm dyn}\equiv\left(\frac{B_{\phi}}{\partial_t^2B_{\phi}}\right)^{1/2}\,.
\end{equation}
The time derivative of the radial magnetic field is
\begin{equation}
    \partial_tB_r\sim\frac{\omega_{\rm A}^2\delta v_{\rm A}}{N\Omega r}\delta B_{\perp}\sim\frac{\omega_{\rm A}^3\delta v_{\rm A}}{N\Omega^2 r}B_{\phi}\sim\frac{\omega_{\rm A}^5}{N\Omega^3}B_{\phi}\,,
\end{equation}
using Eq.~\eqref{eq:sat_TI}. Therefore,
\begin{equation}
    \partial^2_tB_{\phi}\sim q\Omega\partial_tB_r\sim\frac{q\omega_{\rm A}^5}{N\Omega^2}B_{\phi}\,,
\end{equation}
where $q$ and $\Omega$ are assumed to be constant during this phase. Thus, the dynamo characteristic timescale can be approximated as
\begin{equation}\label{eq:t_dyn}
    \tau_{\rm dyn}\equiv\left(\frac{B_{\phi}}{\partial_t^2B_{\phi}}\right)^{1/2}\sim\frac{\Omega}{\omega_{\rm A,dyn}^2}\left(\frac{N}{q\omega_{\rm A,dyn}}\right)^{1/2}\,.
\end{equation}

For the case $P_{\rm fin}=\SI{10}{ms}$, we have  $\tau_{\rm shear}\simeq\SI{2.3}{s}$, $\tau_{\rm TI}\simeq\SI{8.5}{s}$, and $\tau_{\rm dyn}\simeq\SI{5}{s}$, which are similar to the timescales illustrated by the dotted vertical lines in Fig.~\ref{fig:typical_case}. The same observation can be made for $P_{\rm fin}\leq\SI{30}{ms}$ in Fig.~\ref{fig:diff_mass}. However, for $P_{\rm fin}=\SI{40}{ms}$ 
($M_{\rm acc}=\SI{0.008}{M_\odot}$)
and $P_{\rm fin}=\SI{60}{ms}$ 
($M_{\rm acc}=\SI{0.0054}{M_\odot}$),
the dynamo loop phase lasts respectively $\sim\SI{30}{s}$ and $\sim\SI{20}{s}$ (see Fig.~\ref{fig:diff_mass}), which is longer than the analytical predictions of $\tau_{\rm dyn}\simeq\SI{11}{s}$ and $\tau_{\rm dyn}$~$\simeq$~$\SI{9.2}{s}$. This is due to the presence of a significant stage that is not included in the expression of $\tau_{\rm dyn}$ where the growth of $B_{\phi}$ slows down before saturation.
\begin{figure}[tbp]
   \centering
   \includegraphics[width=0.49\textwidth]{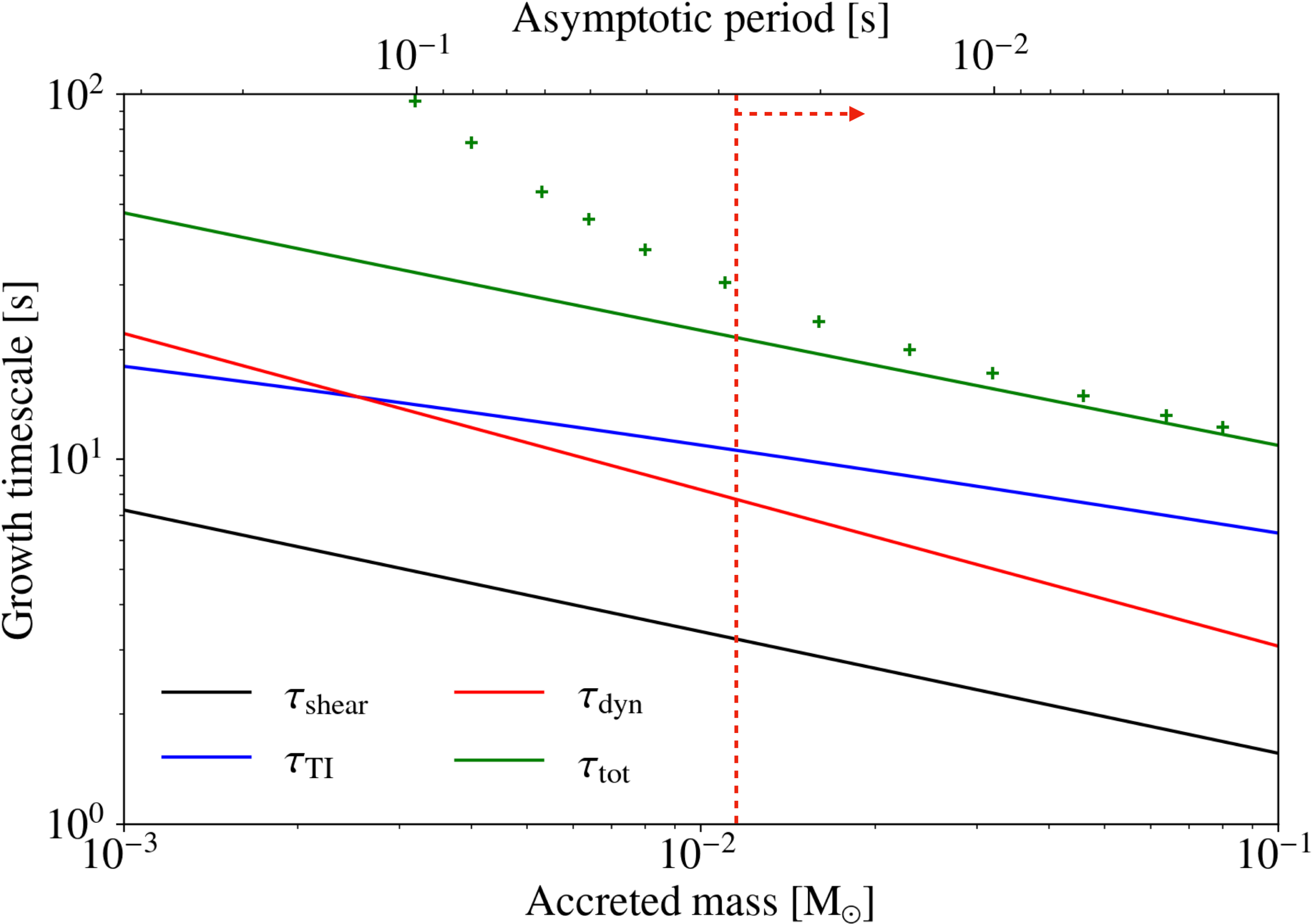}
   \caption{Different characteristic timescales as a function of the accreted mass: winding (black), Tayler instability (dark blue), and dynamo (red). The green line represents the sum of the three timescales. The shear rate is set at $q=1$. The green crosses represent the entire amplification time obtained by integrating Eqs.~\eqref{eq:Bphi_sat}--\eqref{eq:Br_sat}. The red vertical line shows the lower limit of the accreted mass to form a magnetar with a radial field stronger than $B_{\rm Q} \equiv m_{e}c^2/e\hbar\simeq\SI{4.4e13}{G}$ using the predictions of \citet{fuller2019}.}\label{fig:timescales}
\end{figure}

The three characteristic timescales defined by Eqs.~\eqref{eq:t_shear}, \eqref{eq:t_TI}, and \eqref{eq:t_dyn} are plotted as a function of the fallback mass in Fig.~\ref{fig:timescales} in addition to the characteristic timescale for the whole amplification process, which is defined as
\begin{equation}
    \tau_{\rm tot}\equiv\tau_{\rm shear}+\tau_{\rm TI}+\tau_{\rm dyn}\,.
\end{equation}
The vertical red dashed line represents the lower limit of fallback mass to form typical magnetars, which is estimated in Sect.~\ref{ssec:sat_reg} (corresponding to an asymptotic rotation period ${P_{\rm fin}\lesssim \SI{30}{ms}}$). In the regime relevant to magnetar formation, the analytical and numerical estimates of the duration of the whole amplification process are in reasonable agreement, namely  $\tau_{\rm tot}\lesssim\SI{30}{s}$. In this regime, the phase which takes more time is the development of the Tayler instability. For $P_{\rm fin}\gtrsim\SI{30}{ms}$, the comparison between the time at which $B_{\phi}$ saturates and $\tau_{\rm tot}$ shows a significant difference, which is the consequence of the discrepancy noted above between the analytical estimate and numerical solution for $\tau_{\rm dyn}$.


\subsection{Magnetic field in the saturated regime}\label{ssec:sat_reg}

We now focus on the maximum magnetic field obtained at the end of the amplification phase. In the following discussion, this saturated magnetic field will be considered as a proxy for the magnetar's magnetic field and its `radial' component will be considered a proxy for the `dipolar' component of the magnetic field. A more precise prediction would require a description of the relaxation towards a stable equilibrium, which is left for future studies.

In the top panel of Figs.~\ref{fig:typical_case} and \ref{fig:diff_mass}, we see that the saturation intensities are close to their associated horizontal dashed lines, which illustrate the predictions of Eqs.~\eqref{eq:Bphi_sat}--\eqref{eq:Br_sat} for values of the shear rate $q$ and the angular rotation frequency $\Omega$ reached at the time of magnetic field saturation. Therefore, these equations can be used to estimate the intensity of the saturated magnetic field. However, the angular frequency at $\tau_{\rm tot}$ is still lower than its asymptotic value represented by the violet dashed line in the bottom panel. We estimate $\Omega(t=\tau_{\rm tot})$ analytically by integrating Eq.~\eqref{eq:omg_evol} 
\begin{equation}
    \Omega(t=\tau_{\rm tot})=\Omega_{\rm fin}-\left(\frac{t_0}{\tau_{\rm tot}+t_0}\right)^{2/3}(\Omega_{\rm fin}-\Omega_{\rm init})\,.
\end{equation}
Assuming that the timescales for the dynamo are roughly the same for the two models, that is, that of \citet{fuller2019} and that of \citet{spruit2002}, we also evaluate the expressions of the saturated magnetic field derived by \citet{spruit2002} (Eqs. \ref{eq:Bphi_spruit} and \ref{eq:Br_spruit}) at $\Omega(t=\tau_{\rm tot})$.
\begin{figure}[tbp]
   \centering
   \includegraphics[width=0.49\textwidth]{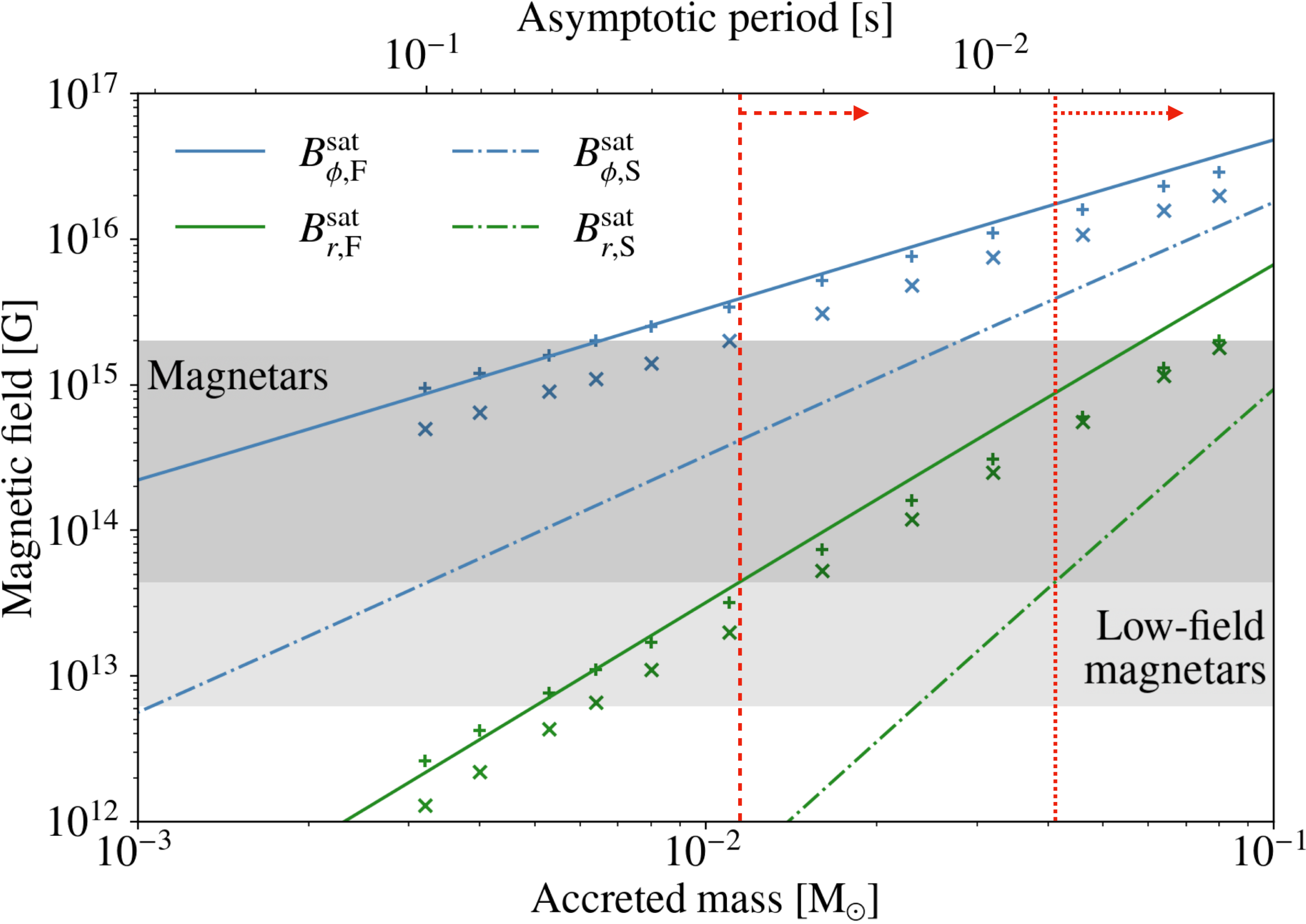}
   \caption{Predicted intensities for the saturated components of the magnetic field as a function of the accreted mass using the formalisms of \citet{fuller2019} (Eqs.~\eqref{eq:Bphi_sat}--\eqref{eq:Br_sat}; solid lines) and \citet{spruit2002} (Eqs.~\eqref{eq:Bphi_spruit}--\eqref{eq:Br_spruit}; dash-dot lines). The shear rate is set at $q=1$. These intensities are compared to the magnetic field reached at maximum intensity (blue and green plus signs) and at $q=0$ (blue and green cross signs) by integrating Eqs.~\eqref{eq:omg_evol}--\eqref{eq:Br_evol} for several fallback masses. Grey areas represent the estimated range of the dipolar magnetic field strength from regularly observed magnetars (dark grey) and from the three detected low-field Galactic magnetars (light grey). The vertical lines show the lower limit on the fallback mass (upper limit on the rotation period) needed to form a magnetar with a radial field stronger than $B_{\rm Q}\simeq\SI{4.4e13}{G}$ for the prediction of \citet{fuller2019} (dashed red) and \citet{spruit2002} (dotted red).}
              \label{fig:saturated_B}
\end{figure}
We see in Fig.~\ref{fig:saturated_B} that our analytical estimates of the saturated fields (solid lines) are close to the numerical values at the peak of the solutions of Eqs.~\eqref{eq:Bphi_evol}--\eqref{eq:Br_evol} (plus symbols). The small difference that appears for shorter rotation periods is due to the angular momentum transport, as discussed in see Sect.~\ref{ssec:AM}. 

Using the maximum magnetic field as a proxy for the magnetar's magnetic field may lead to an overestimation because a fraction of the magnetic energy can be dissipated during the relaxation to a stable magnetic configuration. Although our model is unable to describe this relaxation process, we can get an idea of the robustness of our proxy by comparing to the magnetic field intensity at the time when $q=0$. Figure~\ref{fig:saturated_B} shows that this other proxy (cross signs) is smaller by between $\sim\SI{10}{\%}$ and $\sim\SI{50}{\%}$ but follows the same trends as the maximum magnetic field. Such a moderate difference would not change our main conclusions and suggests that the maximum magnetic field is a meaningful proxy for the final magnetic field.

In Fig.~\ref{fig:saturated_B}, the observed range of dipolar magnetic field for magnetars is fixed between the quantum electron critical field $B_{\rm Q}\equiv m_{e}c^2/e\hbar\simeq\SI{4.4e13}{G}$ and $B_{\rm dip}\sim\SI{2e15}{G}$, the dipole field of the `most magnetised' magnetar SGR 1806-20  \citep{olausen2014}. We find that the radial magnetic fields $B_{r,\rm F}^{\rm sat}$ and $B_{r,\rm S}^{\rm sat}$ fall in this range for accreted masses $M_{\rm acc}\gtrsim\SI{1.1e-2}{\rm M_{\odot}}$ and $M_{\rm acc}\gtrsim\SI{4e-2}{\rm M_{\odot}}$ (i.e. asymptotic rotation periods $P_{\rm fin}\lesssim\SI{28}{ms}$ and $P_{\rm fin}\lesssim\SI{8}{ms}$), respectively. This confirms that magnetar-like magnetic fields can be formed over a wide range of accreted masses. The analytical predictions also show that, in the regime relevant for magnetar formation, the azimuthal component $B_{\phi}\gtrsim\SI{4e15}{G}$ for both saturation mechanisms, which is significantly stronger than the radial component.

For lower accreted masses spinning-up the PNS to periods ranging from $\SI{28}{}$ to $\SI{64}{ms}$ (between $\SI{8}{}$ and $\SI{14}{ms}$ for the predictions of \citet{spruit2002}), our scenario may produce radial magnetic fields $B_r$ as strong as the dipolar fields diagnosed in low-field magnetars \citep{rea2010,rea2012,rea2014}. Moreover, the strength of the associated azimuthal field is $B_{\phi}\sim(1-3)\times 10^{15}\,\SI{}{G}$, which can be related to the non-dipolar magnetic field needed to produce the outbursts and chaotic bursts observed in magnetars \citep{thompson1995}. This azimuthal magnetic field may also be the source of the proton cyclotron absorption lines observed in two low-field magnetars by \citet{tiengo2013} and \citet{rodriguez2016}.
Thus, our model provides a possible explanation of low-field magnetar formation. It is an alternative to the initial interpretation proposed by \citet{rea2010}, which invokes $\gtrsim\SI{1}{Myr}$ `old' (or `worn-out') magnetars whose initial strong dipolar field of $\sim{\rm few}\times\SI{e14}{G}$ has decayed due to Ohmic and Hall processes. This diffusion could be enhanced by the presence of a strong initial toroidal field $\gtrsim\SI{e16}{G}$ \citep{turolla2011}.

As Eqs.~\eqref{eq:Bphi_sat}--\eqref{eq:Br_sat} give orders of magnitude, \citet{fuller2019} parameterised them with a prefactor denoted $\alpha$. We used $\alpha\sim \SI{1}{}$ as obtained by  \citet{fuller2019} for evolved stars in the subgiant and red giant branches by calibrating $\alpha$ on asteroseismic measurements. However, \citet{eggenberger2019} find $\alpha\sim 0.5$ for subgiant stars on the one hand, and $\alpha\sim1.5$ for red giant stars on the other. Also, \citet{fuller2022} argue that $\alpha\sim\SI{0.25}{}$ if intermittent dynamo action is considered in radiative zones with insufficient shear to trigger a sustained dynamo. This smaller prefactor would imply a larger limit of accreted mass of $\sim\SI{2e-2}{M_{\odot}}$ (i.e. a rotation period of $\sim\SI{15}{ms}$).

\section{Discussion}\label{sec:discu}

\subsection{Angular momentum transport}\label{ssec:AM}

In the previous section, we focused on the magnetic field amplification and did not discuss the angular momentum transport due to the Tayler-Spruit dynamo. Our analytical estimate of the saturated magnetic field is based on the assumption that the differential rotation is not erased before the end of the amplification. Indeed, Figs.~\ref{fig:typical_case} and~\ref{fig:diff_mass} show that the angular momentum transport due to Maxwell stresses starts to be significant around the time of magnetic field saturation and that most of the angular momentum transport occurs afterwards. This can be explained by comparing the characteristic timescales of the dynamo loop phase $\tau_{\rm dyn}$ (Eq.~\ref{eq:t_dyn}) with those of angular momentum transport. 
These can be estimated at saturation using the expression of $B_{\phi,\rm F}^{\rm sat}$ (Eq.~\ref{eq:Bphi_sat}) as
\begin{equation}
    \tau_{\rm AM}\equiv\gamma_{\rm AM}^{-1}=\frac{r^2}{\nu_{\rm AM}}\sim\left(\frac{N}{\Omega}\right)^2\Omega^{-1}\,,
\end{equation}
and
\begin{equation}
    \tau_{\rm dyn} = \left(\frac{N}{q \omega}\right)^{4/3} \Omega^{-1} \,.
\end{equation}
The ratio of these two timescales is 
\begin{equation}
    \frac{\tau_{\rm dyn}}{\tau_{\rm AM}} \sim \left(\frac{\Omega}{q^2N}\right)^{2/3} \sim 0.3 \left(\frac{P}{\SI{10}{ms}}\right)^{-2/3} \,.
\end{equation}
For all values of the accreted mass and corresponding rotation period considered in this paper, the angular momentum transport is therefore longer than the dynamo timescale. This explains why most of the angular momentum takes place after the dynamo saturation and justifies a posteriori our analytical estimate of the saturated magnetic field. At fast rotation periods of a few milliseconds, the two timescales are nonetheless close to each other; as a consequence, the angular momentum transport before saturation is not negligible, which explains the moderate discrepancy between our analytical estimate of the saturated magnetic field and the numerical results for short rotation periods (Fig.~\ref{fig:saturated_B}). 

On the other hand, angular momentum transport due to the neutrino viscosity can be neglected because its typical timescale is much longer than the evolution timescales considered:
\begin{equation}
    \tau_{n}\equiv\frac{r^2}{\nu_{n}}\gtrsim\SI{3e4}{s}\,,
\end{equation}
where the neutrino viscosity $\nu_n$ is estimated with the approximate analytical expression of \citet{keil96} and \citet{guilet2015}: 
\begin{equation}\label{eq:neutrino_visc}
\nu_n \sim \SI{3e8}{} \left(\frac{\rho}{\SI{e14}{g.cm^{-3}}}\right)^{-2} \left(\frac{T}{\SI{5}{MeV}}\right)^2\SI{}{cm^2.s^{-1}}.
\end{equation}

\subsection{Neutrinos}\label{ssec:neutrinos}
We demonstrate above that angular momentum transport by either the magnetic field or the neutrino viscosity does not significantly impact our results. However, the neutrino flux coming from the accretion is also expected to extract a fraction of the angular momentum of the PNS \citep{janka2004,bollig2021}. To investigate whether it does not jeopardise the model, we use the following reasoning. As fallback is assumed to start several seconds after bounce in our model, one may assume that most of the angular momentum extraction by neutrino emission at these late times originate from fallback accretion rather than PNS cooling. Most of the fallback mass is likely to have a specific angular momentum $j_0$ which exceeds the Keplerian value at the PNS surface. It will therefore assemble into an accretion disk, settling at a radius $r_\mathrm{k}$ where $j_\mathrm{kep}(r_\mathrm{k})=j_0$. Its gravitational binding energy $E_\mathrm{bind}$ will be at most all converted into neutrino radiation, that is per baryon with a rest mass $m_\mathrm{B}$:
\begin{equation}
    \Delta E_{\nu}\lesssim E_\mathrm{bind}(r_\mathrm{k})\sim\frac{G{M}_{\rm PNS}{m}_{\rm B}}{ r_\mathrm{k}}\,,
\end{equation}
where we assume that the disk mass is small compared to the PNS mass. The corresponding specific angular momentum loss is
\begin{equation}
    \Delta j_{\nu}\lesssim\frac{\Delta E_{\nu}}{m_\mathrm{B}c^2}j_\mathrm{kep}(r_\mathrm{k})\sim\frac{R_\mathrm{S}}{2r_\mathrm{k}}j_\mathrm{kep}(r_\mathrm{k})\,,
\end{equation}
where $R_\mathrm{S}\equiv 2GM_\mathrm{PNS}/c^2$ is the PNS Schwarzchild radius. $\Delta j_{\nu}$ is maximal at the PNS surface (i.e. when $r_\mathrm{k}=r$), which implies
\begin{equation}
    \Delta j_{\nu}\lesssim 0.185 j_\mathrm{kep}(r)\,,
\end{equation}
for the same parameters of a typical PNS introduced in Sect.~\ref{ssec:gov_eq}. Therefore, the extraction of angular momentum by neutrino radiation is very inefficient.

\subsection{Impact of the viscosity on the Tayler instability}
In the reasoning developed above, we do not take into account the effects due to viscous processes, which might be important because they could be much larger than the effects of the resistivity in PNSs. Therefore, here we aim to address the question of their impact on the development of the Tayler instability. To the best of our knowledge, no analytical study of the Tayler instability has included the impact of viscosity. 
Hence, we use an approximate reasoning  similar to that of \citet{spruit2002}, which is based on a comparison of the instability growth timescale with the viscous damping timescale. This provides the following instability criterion:
\begin{equation}
    \sigma_{\rm TI}^{-1}\sim\frac{\Omega}{\omega_{\rm A}^2}\lesssim \frac{l_r^2}{\nu}\,,
\end{equation}
with $\nu$ being the kinematic viscosity. Using the constraint on the radial length scale $l_r$ due to the stratification (Eq.~\ref{eq:strat}), we infer an instability criterion on the azimuthal magnetic field $B_{\phi}$ as a function of the viscosity:
\begin{equation}\label{eq:Bcrit_visc}
    B_{\phi}>B_{\phi\rm, c} \sim\Omega\left(\frac{N}{\Omega}\right)^{1/2}\left(\frac{\nu}{r^2\Omega}\right)^{1/4}\,.   
\end{equation}

This equation is similar to Eq.~\eqref{eq:Bcrit_tayler} but with the magnetic diffusivity substituted by the viscosity. To obtain an order of magnitude of this critical value, we must determine a value of the viscosity which is relevant for our scenario. As the fallback accretion occurs seconds to minutes after the PNS formation, the PNS has cooled down to temperatures $\lesssim\SI{1.1e11}{K}$ in the core and $\lesssim\SI{5e10}{K}$ in the outer region \citep[$\lesssim\SI{10}{MeV}$ and $\lesssim\SI{5}{MeV}$, respectively;][]{hudepohl2014}. The neutrino mean free path can be approximated by \citep[][Eq. 11]{thompson1993}:
\begin{equation}
    l_{n}\sim\SI{4e4}{}\left(\frac{\rho}{\SI{e14}{g.cm^{-3}}}\right)^{-1/3}\left(\frac{T}{\SI{5}{MeV}}\right)^{-3}\left(\frac{f(Y_p)}{1}\right)\SI{}{cm}\,,
\end{equation}
where $f(Y_p)$ is function of the proton fraction close to unity. This length is larger than the maximum radial length scale (Eq.~\ref{eq:strat}):
\begin{equation}
    l_r\sim \SI{4e3}{}\left(\frac{B_{\phi}}{\SI{e15}{G}}\right)\SI{}{cm}\,.
\end{equation}
Therefore, neutrinos do not provide any relevant viscosity at the Tayler instability length scales and we must consider instead a microscopic viscosity such as the shear viscosity due to neutron--neutron scattering \citep[][Eq. 14]{cutler1987}:
\begin{equation}
    \nu_{s}\sim\SI{3e-2}{}\left(\frac{\rho}{\SI{e14}{g.cm^{-3}}}\right)^{5/4}\left(\frac{T}{\SI{5}{MeV}}\right)^{-2}\SI{}{cm^2.s^{-1}}\,.
\end{equation}
The associated critical magnetic field is therefore
\begin{equation}
    B_{\phi\rm, c}\simeq\SI{e13}{}\left(\frac{\nu_s}{\SI{3e-2}{cm^2.s^{-1}}}\right)^{1/4}\left(\frac{\Omega}{\SI{200\pi}{rad.s^{-1}}}\right)^{1/4}\SI{}{G}\,,
\end{equation}
which is four times stronger than the critical magnetic field inferred from the criterion of \citet{spruit2002} (Eq.~\ref{eq:Bcrit_tayler}). However, this new critical magnetic field is still much weaker than the characteristic azimuthal magnetic field separating the winding phase from the phase in which the Tayler instability develops (when the growth rate of the Tayler instability reaches the winding rate (Eq.~\ref{eq:omega_TI})) 
\begin{equation}
    B_{\phi}=\omega_{\rm A,TI}\sqrt{4\pi\rho r^2}\sim\SI{1.3e15}{}\left(\frac{\Omega}{\SI{200\pi}{rad.s^{-1}}}\right)^{2/3}\left(\frac{B_r}{\SI{e12}{G}}\right)^{1/3}\,.
\end{equation}
As a consequence, the viscosity is not expected to prevent the Tayler instability from growing and should not have a significant impact on our results. However, we note that our argument is approximate and would need to be upgraded through a linear analysis of the Tayler instability including the viscous processes.

\subsection{Superfluidity and superconductivity}\label{ssec:super}

A last potential obstacle for our model may emerge from the crust formation and the superfluidity and superconductivity in the core, which occur during the cooling of the PNS. The outer crust is expected to start freezing a few minutes after the PNS formation and the inner crust forms far later, between \SI{1}{} and \SI{100}{yr} after formation \citep{Aguilera2008}. Therefore, no part of the crust is formed during the time interval involved in our scenario. 

The potential early appearance of superfluid neutrons or even superconductive protons in the PNS core at temperatures below $\SI{e8}{}-\SI{e10}{K}$ is worth discussing because the MHD approximation would not be sufficiently realistic and a multi-fluid approach would be more relevant \citep{glampedakis2011,sinha2015}. However, the 1D models of PNS cooling show higher temperatures than \SI{e10}{K} in the PNS even after \SI{15}{s} \citep[e.g.][]{pons1999,roberts2012,hudepohl2014,roberts2017}. Moreover, \citet{gusakov2013} and \citet{glampedakis2014} brought forward a critical perturbed magnetic field strength above which superfluidity of neutrons dies out. Therefore, the MHD approximation is still valid for describing the PNS internal dynamics during the first \SI{40}{s} following the core bounce.\\

\section{Conclusions}\label{sec:conclu}

In this paper, we propose a new scenario for magnetar formation, in which the Tayler-Spruit dynamo amplifies the large-scale magnetic field of a PNS spun up by SN fallback accretion. We develop a one-zone model describing the evolution of the magnetic field averaged over a PNS subject to fallback accretion. The equations describing the time evolution are solved numerically and compared successfully with analytical estimates of the final magnetic field and of the duration of each stage of the amplification process. Predictions for the different components of the magnetic field are therefore obtained as a function of the accreted mass for the two proposed saturation models of the Tayler-Spruit dynamo \citep{spruit2002,fuller2019}. Our main conclusions can be summarised as follows:
\begin{itemize}
    \item Radial magnetic fields spanning the full range of the magnetar dipole intensity can be formed for accreted masses compatible with the results of recent SN simulations. Our model predicts the formation of magnetar-like magnetic fields for accreted masses $M_{\rm acc}\gtrsim\SI{1.1e-2}{M_{\odot}}$ for the saturation model of \citet{fuller2019} and $M_{\rm acc}\gtrsim\SI{4e-2}{M_{\odot}}$ for the saturation model of \citet{spruit2002}. This corresponds to neutron star final rotation periods $P_{\rm fin}\lesssim\SI{28}{ms}$ and $P_{\rm fin}\lesssim\SI{8}{ms}$, respectively.
    \item The azimuthal component of the magnetic field is predicted to be in the range $10^{15}-\SI{e16}{G}$, which is stronger than the radial component by a factor of 10 to 100. 
    \item In the regime relevant for magnetar formation, the magnetic field amplification lasts between 15 and $\SI{30}{s}$. On such a timescale, the MHD equations assumed in the description of the Tayler-Spruit dynamo are expected to be valid. Furthermore, we have not identified any other process capable of interfering with the Tayler-Spruit dynamo by transporting angular momentum on a comparable or shorter timescale.
\end{itemize}
Our results therefore predict that magnetars can indeed be formed in our new scenario. Magnetar formation is possible at sufficiently long rotation periods to be compatible with the lower limit of $\SI{5}{ms}$ inferred from regular SN remnants associated with magnetars. With the saturation model of \citet{fuller2019}, the full range of magnetar fields can be obtained within this constraint, even those that exhibit a strong dipolar magnetic field $B_{\rm dip}\sim\SI{e15}{G}$. On the other hand, with the saturation model proposed by \citet{spruit2002}, only the lower end of the magnetar fields can be obtained with $P_{\rm fin}<\SI{5}{ms}$, while dipolar magnetic fields $\gtrsim\SI{2e14}{G}$ need faster rotation periods.

An important prediction of our scenario is the very intense toroidal magnetic field, which lies between $3\times 10^{15}$ and $\SI{3e16}{G}$ for parameters corresponding to radial magnetic fields in the magnetar range. These values are compatible with the interpretation of the X-ray flux modulations observed in three magnetars as free precession driven by an intense toroidal magnetic field \citep{makishima2014,makishima2016,makishima2019,makishima2021}.

The intense toroidal magnetic field predicted in our scenario also provides interesting perspectives from which to explain the formation of low-field magnetars. For radial magnetic fields in the range of the dipolar magnetic field deduced for these objects \citep{rea2010,rea2012,rea2013,rea2014}, our model predicts a toroidal magnetic field intensity of $\sim\SI{1}{}-\SI{3e15}{G}$. Such non-dipolar magnetic fields are strong enough to be the energy source of the magnetar-like emission from these objects and to explain the variable absorption lines interpreted as proton cyclotron lines \citep{tiengo2013,rodriguez2016}. We therefore suggest that some of the low-field magnetars may be born with low dipolar magnetic fields, rather than evolve to this state as assumed in the `worn-out' magnetar scenario.

A question arising from our study is the location of the magnetic field in the PNS, which cannot be captured by our one-zone model. As the shear due to fallback accretion is expected to be strongest in the outer region of the PNS, one may expect the magnetic field to be preferentially located in these outer layers. Such a concentration of the magnetic field near the surface would have interesting consequences for its long-term evolution because the magnetic field may be confined in the crust without significant magnetic flux threading the superconductive core. The long-term evolution of such a crust-confined magnetic field configuration has been thoroughly investigated by numerical simulations \citep[e.g.][]{vigano2013,gourgouliatos2016,pons2019}. By contrast, if the magnetic field is also present in deeper regions, its evolution in the superconductive core and the transition layer with the crust must be taken into account. A few papers studied this evolution in numerical simulations \citep[e.g.][]{henriksson2013,lander2013,ciolfi2013} but these lead to a slower magnetic field evolution that is incompatible with magnetar observations \citep{elfritz2016}. These results would favour initial crust-confined magnetic fields but need to be confirmed in more realistic 3D simulations of the magneto-thermal evolution in the whole neutron star. We also note that the localisation of the magnetic field in our magnetar formation scenario should be studied in more detail. On the one hand, the stratification increases by a factor of $\sim 2.5$ close to the PNS surface, which might weaken the magnetic field and confine it closer to the surface. On the other hand, the shear can be expected to become significant in the bulk of the PNS after angular momentum has been partly redistributed by the Tayler-Spruit dynamo. Some of the magnetic field may also be transported to deeper regions via the Tayler instability or during the relaxation to a stable equilibrium.

Another relevant question is the geometry of the magnetic field amplified by the Tayler-Spruit dynamo. One should keep in mind that our comparison to magnetars relies on the assumption that the generated radial field $B_r$ is mostly dipolar. Although the real geometry of the poloidal magnetic field generated by the Tayler-Spruit dynamo is not known, it is likely to be partly non-dipolar, meaning that the large-scale dipolar magnetic field is a fraction of the radial field $B_r$. Therefore, corresponding predictions should be refined by studying dedicated multi-dimensional models. In \citet{petitdemange2022}, a dynamo similar to the Tayler-Spruit dynamo has been found through numerical simulations in a configuration where the surface rotates slower than the core, which is therefore different to the case of spun-up PNS. Moreover, the observed magnetars are cooled-down neutron stars with a stable configuration of magnetic field. Hence, the study of the magnetic field relaxation from a turbulent saturated state to a stable configuration is important to estimate a more realistic intensity of the dipolar poloidal field.
Thus, numerical simulations will be essential to further study of the evolution of the magnetic field geometry in our framework.

A salient feature of our fallback scenario is that it decouples magnetar formation from rapid progenitor rotation and from strong magnetisation of the pre-collapse stars. Rapid progenitor rotation is necessary for magnetar formation by the convective dynamo, which requires initial NS spin periods of $\lesssim\SI{10}{ms}$ \citep{raynaud2022}, and by the magnetorotational instability \citep{reboul2021a,reboul2021b}. Strong magnetisation of the pre-collapse star on the other hand is a crucial aspect in the fossil field scenario or the stellar merger scenario \citep{schneider2019}. Instead of requiring fast rotation or strong magnetic field in the progenitor core, our scenario predicts magnetar formation when fallback deposits a sufficient amount of angular momentum on the PNS surface. With the angular momentum of the mass accreted by the NS being limited by the Keplerian value, magnetars are formed for accreted masses of more than $\sim\SI{1.1e-2}{{\rm M}_{\odot}}$ \citep[case of][]{fuller2019} and $\sim\SI{4e-2}{{\rm M}_{\odot}}$ \citep[case of][]{spruit2002} in our scenario. The fallback mass should be several times larger than the accreted mass, because angular momentum loss must be expected to lead to mass loss during the accretion process. Therefore, fallback masses of more than a few $\SI{e-2}{{\rm M}_{\odot}}$ to $\SI{e-1}{{\rm M}_{\odot}}$ seem to be needed. Based on 1D models of neutrino-driven core-collapse SN explosions, this indicates a preference for single stars with zero-age-main-sequence (ZAMS) masses above about $\sim\SI{18}{{\rm M}_{\odot}}$ \citep{sukhbold2016} and helium stars (hydrogen-stripped stars in binaries) with ZAMS masses above $\SI{30}{}-\SI{40}{{\rm M}_{\odot}}$ (depending on details of the mass-loss evolution); although the compactness differences between the single-star models of \citet{sukhbold2014} compared to those of \citet{sukhbold2018} as well as 3D explosion effects \citep[which increase the fallback mass;][]{janka2021} may shift these ZAMS masses to lower values. This would be consistent with the observations constraining magnetar progenitors to masses higher than $\SI{30}{M_{\odot}}$ \citep{gaensler05,bibby08,clark08} and also with the case of the magnetar SGR1900+14, whose progenitor mass was estimated to be $\SI{17}{}\pm\SI{2}{M_{\odot}}$ \citep{davies2009}.

While our model avoids the uncertainty of the progenitor core rotation and magnetic field, it implies coping with the uncertainties on the fallback process. A precise modelling of the fallback depends on such challenging questions as how a long-lasting post-explosion phase where downflows to the PNS coexist with outflows of neutrino-heated matter transitions into the fallback accretion as discussed by \citet{janka2021}; the complex dynamical processes that determine the fraction of fallback matter that gets accreted by the PNS from a fallback disk; and the efficiency of the accretion to spin-up the PNS. Our scenario should therefore be explored in more depth by more realistic fallback models.

Following its saturation, the PNS magnetic field may interact with the newly formed disk of fallback matter and is strong enough to influence the fallback accretion mechanism. We did not model this interaction because our study was focused on the phase of magnetic field amplification. Nevertheless, this could strongly influence the rotation of the newly born magnetar. The evolution of the PNS-fallback disk system depends on three characteristic radii \citep{metzger2018,beniamini2020,lin2020,ronchi2022}: (i) the magnetospheric radius~$r_{\rm m}$, which is the radius at which the matter is blocked by the magnetic barrier, (ii) the corotation radius~$r_{\rm c}$ where the matter has the same rotation frequency as the PNS, and (iii) the light cylinder radius~$r_{\rm lc}$, which is the ratio of light speed to the PNS rotation frequency. The strong magnetic field repels the magnetosphere behind the corotation radius (i.e. $r_{\rm c}<r_{\rm m}$) which stops the accretion and so the PNS spin-up. If the fallback accretion rate of the disk is large enough, the inner part of the disk penetrates the light cylinder (i.e. $r_{\rm lc}>r_{\rm m}$) and opens up a part of the magnetic field lines. The PNS-fallback disk system enters the so-called propeller regime and the PNS angular momentum is transported towards the disk via the magnetic dipole torque. This mechanism is thought to extract the PNS angular momentum very efficiently; for instance \citet{beniamini2020} even predict magnetars spun down to rotation periods of $\sim\SI{e6}{s}$ after $\sim\SI{e3}{yr}$. For this reason, this scenario is often invoked to explain the ultra-long-period magnetars such as 1E 1613 \citep[e.g.][]{deluca2006,li2007,rea2016} or the recently observed GLEAM-X J162759.5-523504.3 \citep{ronchi2022}, which have respective rotation periods of $\sim\SI{2.4e4}{s}$ and $\sim\SI{1.1e3}{s}$. It would be interesting to include such a spin-down model in our magnetar formation scenario in order to obtain a prediction of the rotation period at later times.

Finally, the PNS-fallback disk system has also been invoked to explain the light curve of luminous and extreme SNe of types Ib/c \citep[e.g.][]{dexter2013,metzger2018,lin2021}. We may also expect our scenario to produce these types of explosions depending on the amount of accreted mass during the dynamo process. First, PNSs that have accreted $\sim\SI{2}{}-\SI{3e-2}{M_{\odot}}$ of fallback matter before the magnetic field saturation have rotation periods of around $\SI{10}{}-\SI{20}{ms}$, which are too slow to produce extreme explosions. According to our scenario, their typical magnetic field is of $\SI{1}{}-\SI{5e14}{G}$, which would lead to regular luminous SNe Ib/c. Their light curve would be dominated by the PNS spin-down luminosity instead of the $^{56}$Ni decay luminosity \citep{ertl2020,afsariardchi21}. Second, for fallback masses spinning up the PNSs to millisecond rotation periods, the magnetic field saturates a few $\SI{10}{s}$ after the core bounce at $B_r\gtrsim\SI{5e14}{G}$. The rotational energy can be kept for later times and be slowly extracted to irradiate its environment, which might lead to superluminous SNe I \citep{woosley2010,kasen2010,bersten2016,margalit2018,lin2020,lin2021}. Finally, to produce extreme explosions such as hypernovae, which have approximately ten times larger kinetic energies and much higher $^{56}$Ni yields than the vast majority of CCSNe, an energy injection within a timescale of $\lesssim\SI{1}{s}$ is required \citep{barnes2018} to explain the large masses of $^{56}$Ni $\gtrsim\SI{0.2}{M_{\odot}}$ inferred from their light curves \citep[e.g.][]{woosley2006,drout2011,nomoto2011}. For rotation periods of $\lesssim\SI{1}{ms}$, which correspond to rotational energies of $\gtrsim\SI{3e52}{erg}$, our model provides a radial magnetic field of $B_r\gtrsim\SI{2.6e16}{G}$, which may be enough to inject the energy quickly through magnetic dipole spin-down only \citep{suwa2015}. The presence of a propeller regime would enhance the PNS spin-down such that a weaker dipolar magnetic field of $\sim\SI{2e15}{G}$ would also produce a hypernova \citep{metzger2018} but through a propeller-powered explosion. Therefore, our magnetic-field-amplification scenario by PNS accretion or fallback accretion may be of relevance to a wide variety of magnetar-powered phenomena in different types of SN events.


\begin{acknowledgements}
    We thank the referee Jim Fuller for his thorough reading and relevant comments, which helped improve this paper. This research was supported by the European Research Council through the ERC starting grant MagBURST No. 715368.
    Hans-Thomas Janka acknowledges support by the Deutsche Forschungsgemeinschaft (DFG, German
    Research Foundation) through Sonderforschungsbereich (Collaborative Research Centre) 
    SFB-1258 `Neutrinos and Dark Matter in Astro- and Particle Physics (NDM)' and under
    Germany’s Excellence Strategy through Cluster of Excellence ORIGINS (EXC-2094)—390783311.
    We are grateful to Thierry Foglizzo and Ludovic Petitdemange for useful discussions.
\end{acknowledgements}


%
%
\bibliographystyle{aa}
\bibliography{biblio}

\end{document}